\documentclass[english]{IEEEtran}
\usepackage[T1]{fontenc}
\usepackage[latin9]{inputenc}
\usepackage{color}
\usepackage{array}
\usepackage{multirow}
\usepackage{fixltx2e}
\usepackage{graphicx}
\usepackage{cite}
\usepackage{amsfonts} 
\usepackage{amsmath}

\makeatletter

\newcommand{\noun}[1]{\textsc{#1}}
\providecommand{\tabularnewline}{\\}

\makeatother

\begin{document}

\title{Markov Decision Processes with Applications in Wireless Sensor Networks: A Survey}

\author{\IEEEauthorblockN{ Mohammad Abu Alsheikh\IEEEauthorrefmark{1}\IEEEauthorrefmark{2}, Dinh Thai Hoang\IEEEauthorrefmark{1}, Dusit Niyato\IEEEauthorrefmark{1}, Hwee-Pink Tan\IEEEauthorrefmark{2} and Shaowei Lin\IEEEauthorrefmark{2}}

\IEEEauthorblockA{\IEEEauthorrefmark{1}School of Computer Engineering, Nanyang Technological University, Singapore 639798}

\IEEEauthorblockA{\IEEEauthorrefmark{2}Sense and Sense-abilities Programme, Institute for Infocomm Research, Singapore 138632} }
\maketitle
\begin{abstract}
Wireless sensor networks (WSNs) consist of autonomous and resource-limited devices. The devices cooperate to monitor one or more physical phenomena within an area of interest. WSNs operate as stochastic systems because of randomness in the monitored environments. For long service time and low maintenance cost, WSNs require adaptive and robust methods to address data exchange, topology formulation, resource and power optimization, sensing coverage and object detection, and security challenges. In these problems, sensor nodes are to make optimized decisions from a set of accessible strategies to achieve design goals. This survey reviews numerous applications of the Markov decision process (MDP) framework, a powerful decision-making tool to develop adaptive algorithms and protocols for WSNs. Furthermore, various solution methods are discussed and compared to serve as a guide for using MDPs in WSNs.\end{abstract}


\section{Introduction}

Recent demand for wireless sensor networks (WSNs), e.g., in smart cities, introduces the need for sensing systems that can interact with the surrounding environment's dynamics and objects. However, this interaction is constrained by the limited resources of battery-powered sensor nodes. In many applications, sensor nodes are designed to operate for several months or a few years without battery maintenance~\cite{wu2005lightweight}. The emerging applications of WSNs introduce more resource intensive operations with low maintenance cost requirements. Therefore, adaptive and energy efficient algorithms are becoming more highly valued than ever.

WSNs operate in stochastic (random) environments under uncertainty. In particular, a sensor node, as a decision maker or agent, applies an action to its environment, and then transits from a state to another. The environment can encompass the node's own properties (e.g., location coordinate and available energy in the battery) as well as many of the surrounding objects (e.g., other nodes in the network or a moving target). Thus, the actions can be simple tasks (e.g., switching the radio transceiver into sleep mode to conserve energy), or complex commands (e.g., the moving strategies of a mobile node to achieve area coverage). In such an uncertain environment, the system dynamics can be modeled using a mathematical framework called Markov decision processes (MDPs) to optimize the network's desired objectives. MDPs entail that the system possesses a Markov property. In particular, the future system state is dependent only on the current state but not the past states. Recent developments in MDP solvers have enabled the solution for large scale systems, and have introduced new research potentials in WSNs.

MDP modeling provides the following general benefits to WSNs' operations: 
\begin{enumerate} 
\item WSNs consist of resource-limited devices. Static decision commands may lead to inefficient energy usage. For example, a node sending data at fixed transmit power without considering the channel conditions will drain its energy faster than the one that adaptively manages its transmit power~\cite{udenze2008partially,kobbane2012dynamic}. Therefore, using MDPs for dynamically optimizing the network operations to fit the physical conditions results in significantly improved resource utilization. 
\item The MDP model allows a balanced design of different objectives, for example, minimizing energy consumption and maximizing sensing coverage. Different works, e.g.,~\cite{chatterjee2006markov,etessami2007multi,chatterjee2007markov}, discuss the approaches of using MDPs in optimization problems with multiple objectives. 
\item New applications of WSNs interact with mobile entities that significantly increase the system dynamics. For example, using a mobile gateway for data collection introduces many design challenges~\cite{di2011data}. Here, the MDP method can explore the temporal correlation of moving objects and predicting their future locations, e.g.,~\cite{zhan2010active,atia2011sensor}. 
\item The solution of an MDP model, referred to as a \emph{policy}, can be implemented based on a look-up table. This table can be stored in sensor node's memory for online operations with minimal complexity. Therefore, the MDP model can be applied even for tiny and resource-limited nodes without any high computation requirements. Moreover, near-optimal solutions can be derived to approximate optimal decision policies which enables the design of WSN algorithms with less computation burdens. 
\item MDPs are flexible with many variants that can fit the distinct conditions in WSN applications. For example, sensor nodes generally produce noisy readings, therefore hampering the decision making process. With such imprecise observations, one of the MDP's variants, i.e., partially observable Markov decision process (POMDP), can be applied to reach the best operational policy. Another example of the MDP's flexibility is the use of hierarchical Markov decision process (HMDP) for a hierarchical topology of nodes, cluster heads, and gateways found in WSNs, e.g.,~\cite{yeow2007energy}. 
\end{enumerate}

\begin{figure*}[t]
\begin{centering}
\includegraphics[width=0.7\paperwidth,trim=1cm 0.5cm 1cm 0.5cm]{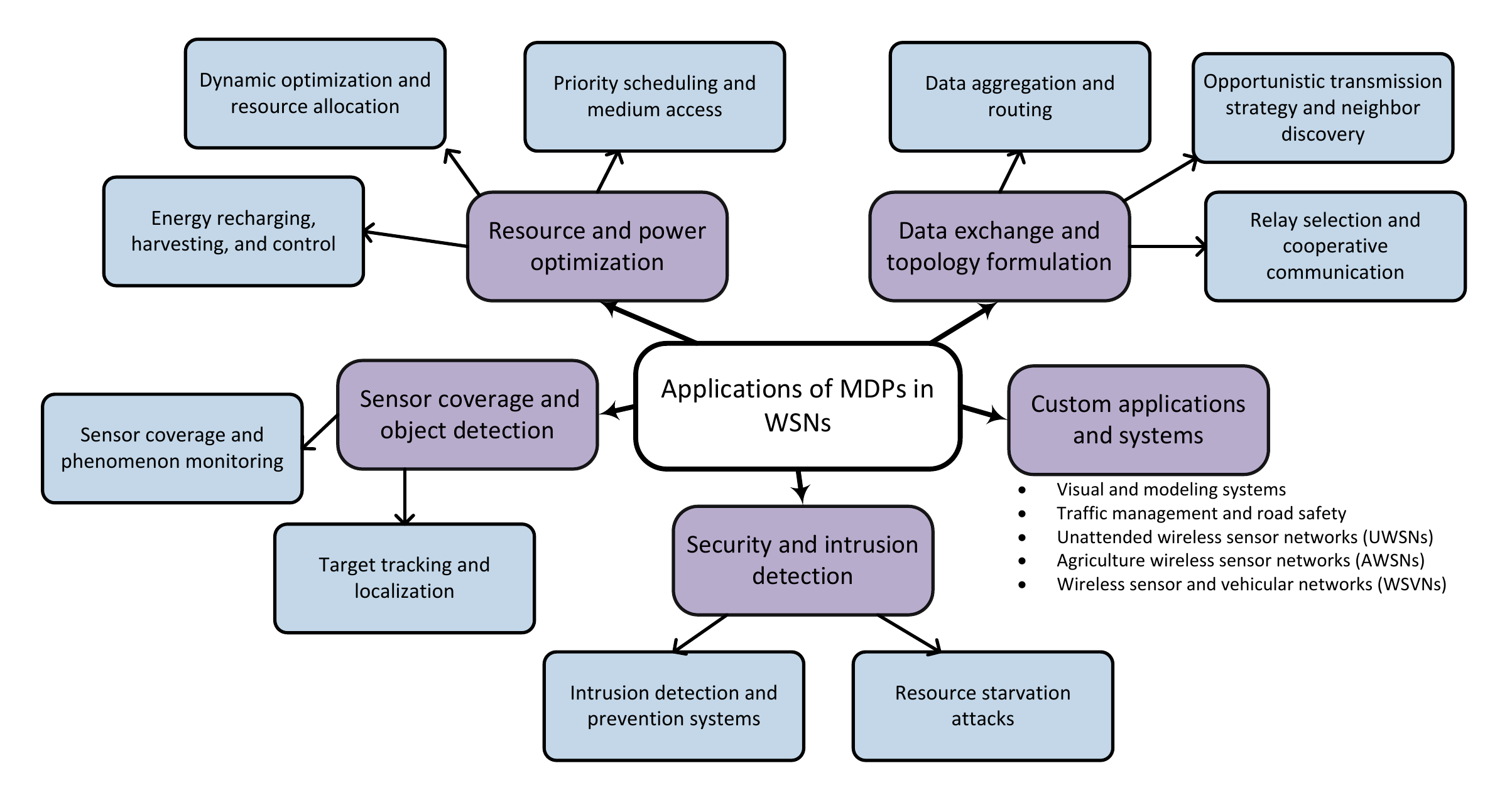}
\par\end{centering}
\caption{\label{fig:taxonomy}Taxonomy of the applications of MDPs in WSNs.}
\end{figure*}

In this paper, we survey the MDP models proposed for solving various design and resource management issues in WSNs. In particular, we classify the related work based on the WSN's issues as shown in Figure~\ref{fig:taxonomy}. The issues include data exchange and topology formation methods, resource and power optimization perspectives, sensing coverage and event tracking solutions, and security and intrusion detection methods. We also review efficient algorithms, which consider the tradeoff between energy consumption and solution optimality in WSNs. Throughout the paper, we highlight the advantages and disadvantages of the solution methods.

Although there are many applications of Markov chains in WSNs, such as data aggregation and routing~\cite{marin2008tandem,ko2010epidemic}, duty cycle~\cite{lee2008enabling}, sensing coverage~\cite{ma2012stochastic}, target tracking~\cite{liu2006cross,zhang2008performance,zheng2012decentralized}, MAC backoff operation~\cite{park2009generalized,sikandar2014optimizing}, and security~\cite{paschalidis2008anomaly,di2009data,chen2014novel}, this paper focuses only on the applications of MDPs in WSNs. The main difference between an MDP and a Markov chain is that the Markov chain does not consider actions and rewards. Therefore, it is used only for performance analysis. By contrast, the MDP is used for stochastic optimization, i.e., to obtain the best actions to be taken given  particular objectives and possibly a set of constraints. The survey on the applications of the Markov chain with WSNs is beyond the scope of this paper.

The rest of this paper is organized as follows. In Section~\ref{sec:markov-decision-process}, a comprehensive discussion of the MDP framework and its solution methods is presented. Then, Sections~\ref{sec:Data-exchange}-\ref{sec:Custom-applications} discuss the applications of MDPs in WSNs. In each section, a problem is first presented and motivated. Then notable studies from the literature are reviewed. Future directions and open research problems are presented in Section~\ref{sec:Future-research}. Finally, the paper is concluded and summarized in Section~\ref{sec:Summary}.

\section{Markov decision processes}\label{sec:markov-decision-process}

A Markov decision process (MDP) is an optimization model for decision making under uncertainty~\cite{Bellman_book_1957_Dynamic_Programming, Puterman_book_1994_Markov_decision}. The MDP describes a stochastic decision process of an agent interacting with an environment or system. At each decision time, the system stays in a certain state $s$ and the agent chooses an action $a$ that is available at this state. After the action is performed, the agent receives an immediate reward $R$ and the system transits to a new state $s'$ according to the transition probability $P_{s,s'}^{a}$. For WSNs, the MDP is used to model the interaction between a wireless sensor node (i.e., an agent) and their surrounding environment (i.e., a system) to achieve some objectives. For example, the MDP can optimize an energy control or a routing decision in WSNs.

\subsection{The Markov Decision Process Framework}

The MDP is defined by a tuple $\langle \mathcal{S}, \mathcal{A}, \mathcal{P}, \mathcal{R}, \mathcal{T} 	\rangle$ where,
\begin{itemize}
\item $\mathcal{S}$ is a finite set of states,
\item $\mathcal{A}$ is a finite set of actions,
\item $\mathcal{P}$ is a transition probability function from state $s$ to state $s'$ after action $a$ is taken,
\item $\mathcal{R}$ is the immediate reward obtained after action $a$ is made, and 
\item $\mathcal{T}$ is the set of decision epoch, which can be finite or infinite.
\end{itemize}

$\pi$ denotes a ``policy'' which is a mapping from a state to an action. The goal of an MDP is to find an optimal policy to maximize or minimize a certain objective function. An MDP can be finite or infinite time horizon. For the finite time horizon MDP, an optimal policy $\pi^*$ to maximize the expected total reward is defined as follows:
\begin{equation}
\label{eq:finite_horizon_MDP_reward_criterion}
\max \mathcal{V}_\pi(s) =	\mathbb{E}_{\pi,s} \bigg[		\sum_{t=1}^{T} \mathcal{R}\Big(s'_t|s_t,\pi(a_t) \Big)	\bigg]	
\end{equation}
where $s_t$ and $a_t$ are the state and action at time $t$, respectively.

For the infinite time horizon MDP, the objective can be to maximize the expected discounted total reward or to maximize the average reward. The former is defined as follows:
\begin{equation}
\max \mathcal{V}_\pi(s) =	\mathbb{E}_{\pi,s} \bigg[		\sum_{t=1}^{T} \gamma^t \mathcal{R}\Big(s'_t|s_t,\pi(a_t) \Big)	\bigg]	,
\end{equation}
while the latter is expressed as follows:
\begin{equation}
\max \mathcal{V}_\pi(s) =	\lim \inf_{T \rightarrow \infty} \frac{1}{T}	\mathbb{E}_{\pi,s} \bigg[		\sum_{t=1}^{T} \mathcal{R} \Big(s'_t|s_t,\pi(a_t) \Big) \bigg]		.
\end{equation}
Here, $\gamma$ is the discounting factor and $\mathbb{E}[\cdot]$ is the expectation function. 

\subsection{Solutions of MDPs}

Here we introduce solution methods for MDPs with discounted total reward. The algorithms for MDPs with average reward can be found in~\cite{Puterman_book_1994_Markov_decision}. 

\subsubsection{Solutions for Finite Time Horizon Markov Decision Processes}
In a finite time horizon MDP, the system operation takes place in a known period of time. In particular, the system starts at state $s_0$ and continues to operate in the next $T$ periods. The optimal policy $\pi^*$ is to maximize $\mathcal{V}_{\pi}(s)$ in~(\ref{eq:finite_horizon_MDP_reward_criterion}). If we denote $v^*(s)$ as the maximum achievable reward at state $s$, then we can find $v^*(s)$ at every state recursively by solving the following \emph{Bellman's optimal equations}~\cite{Bellman_book_1957_Dynamic_Programming}:
\begin{equation}
v_{t}^{*}(s) = \max_{a \in \mathcal{A}}	\bigg[	\mathcal{R}_{t}(s,a)	+	\sum_{s' \in \mathcal{S}}	\mathcal{P}_{t}(s'|s,a)	v^*_{t+1}(s')	\bigg]	.
\end{equation}
Based on the optimal Bellman equations, two typical approaches for finite time horizon MDPs exist.

\begin{itemize}
\item \emph{Backwards induction}: Also known as a dynamic programming approach, it is the most popular and efficient method for solving the Bellman's equations. Since the process will be stopped at a known period, we can first determine the optimal action and the optimal value function at the last time period. We then recursively obtain the optimal actions for earlier periods back to the first period based on the Bellman optimal equations. 

\item \emph{Forward induction}:
This forward induction method is also known as a value iteration approach. The idea is to divide the optimization problem based on the number of steps to go. In particular, given an optimal policy for $t-1$ time steps to go, we calculate the Q-values for $k$ steps to go. After that, we can obtain the optimal policy based on the following equations:
\begin{displaymath} 
\begin{aligned}
&\phantom{1}		Q_t(s,a) = \mathcal{R}(s,a,s') +	\sum_{s'} \mathcal{P}(s,a,s') 	 v^*_{t-1}(s')	,	\\
&\phantom{1}	 	v^*_{t}(s) = \max_{a \in \mathcal{A}} Q_t^*(s,a) 	\phantom{5}  	\text{and}	\phantom{5} 	\pi_t^*(s) = \arg \max_{a \in \mathcal{A}} Q_t^*(s,a)	,
\end{aligned}
\end{displaymath}
where $v_t(s)$ is the value of state $s$ and $Q_t(s,a)$ is the value of taking action $a$ at state $s$. This process will be performed until the last period is reached. 
\end{itemize}

Both approaches have the same complexity which depends on the time horizon of an MDP. However, they are used differently. Backward induction is especially useful when we know the state of MDPs in the last period. By contrast, forward induction is applied when we only know the initial state.

\subsubsection{Solutions for Infinite Time Horizon Markov Decision Processes}

Solving an infinite time horizon MDP is more complex than that of a finite time horizon MDP. However, the infinite time horizon MDP is more widely used because in practice the operation time of systems is often unknown and assumed to be infinite. Many solution methods were proposed.

\begin{itemize}
\item \emph{Value iteration (VI)}: This is the most efficiently and widely used method to solve an infinite time horizon discounted MDP. This method has many advantages, e.g., quick convergence, ease of implementation, and is especially a very useful tool when the state space of MDPs is very large. Similar to the forward induction method of a finite time horizon MDP, this approach was also developed based on dynamic programming. However, for infinite time horizon MDP, since the time horizon is infinite, instead of running the algorithm for the whole time horizon, we have to use a stopping criterion (e.g., $||v^*_{t}(s) - v^*_{t-1}(s)||<\epsilon$) to guarantee the convergence~\cite{Bellman_book_1957_Dynamic_Programming}.

\item \emph{Policy iteration (PI)}: 
The main idea of this method is to generate an improving sequence of policies. It starts with an arbitrary policy and updates the policy until it converges. This approach consists of two main steps, namely policy evaluation and policy improvement. We first solve the linear equations to find the expected discounted reward under the policy $\pi$ and then choose the improving decision policy for each state. Compared with the value iteration method, this method may take fewer iterations to converge. However, each iteration takes more time than that of the value iteration method because the policy iteration method requires solving linear equations.

\item \emph{Linear programming (LP)}: Unlike the previous methods, the linear programming method aims to find a static policy through solving a linear program~\cite{Epenoux_1963_A_probabilistic}. After the linear program is solved, we can obtain the optimal value $v^*(s)$, based on which we can determine the optimal policy $\pi^*(s)$ at each state. The linear programming method is relatively inefficient compared with the value and policy iteration methods when the state space is large. However, the linear programming method is useful for MDPs with constraints since the constraints can be included as linear equations in the linear program~\cite{Altman_book_1995_Constrained_Markov}.

\item \emph{Approximation method}: Approximate dynamic programming~\cite{Warren_book_2011_Approximate_Dynamic} was developed for large MDPs. The method approximates the value functions (whether policy functions or value functions) by assuming that these functions can be characterized by a reasonable number of parameters. Thus, we can seek the optimal parameter values to obtain the best approximation, e.g., as given in~\cite{Sutton_book_1998_Introduction_to, Warren_book_2011_Approximate_Dynamic} and~\cite{Farias_2003_The_linear}.

\item \emph{Online learning}: The aforementioned methods are performed in an offline fashion (i.e., when the transition probability function is provided). However, they cannot be used if the information of such functions is unknown. Learning algorithms were proposed to address this problem~\cite{Sutton_book_1998_Introduction_to, Sigaud_book_2010_Markov_Decision}. The idea is based on the simulation-based method that evaluates the interaction between an agent and system. Then, the agent can adjust its behavior to achieve its goal (e.g., trial and error). 

\end{itemize}

Note that the solution methods for discrete time MDPs can be applied for continuous time MDPs through using uniformization techniques~\cite{Gallager_book_1995_Discrete_Stochastic, Xianping_book_2009_Introduction_to}. The solutions of discrete time MDPs that solve the continuous time MDPs are also known as \emph{semi-MDPs} (SMDPs).

\subsection{Extensions of MDPs and Complexity}\label{sec:mdp_complexity}
\begin{figure}
\begin{centering}
\includegraphics[scale=0.55]{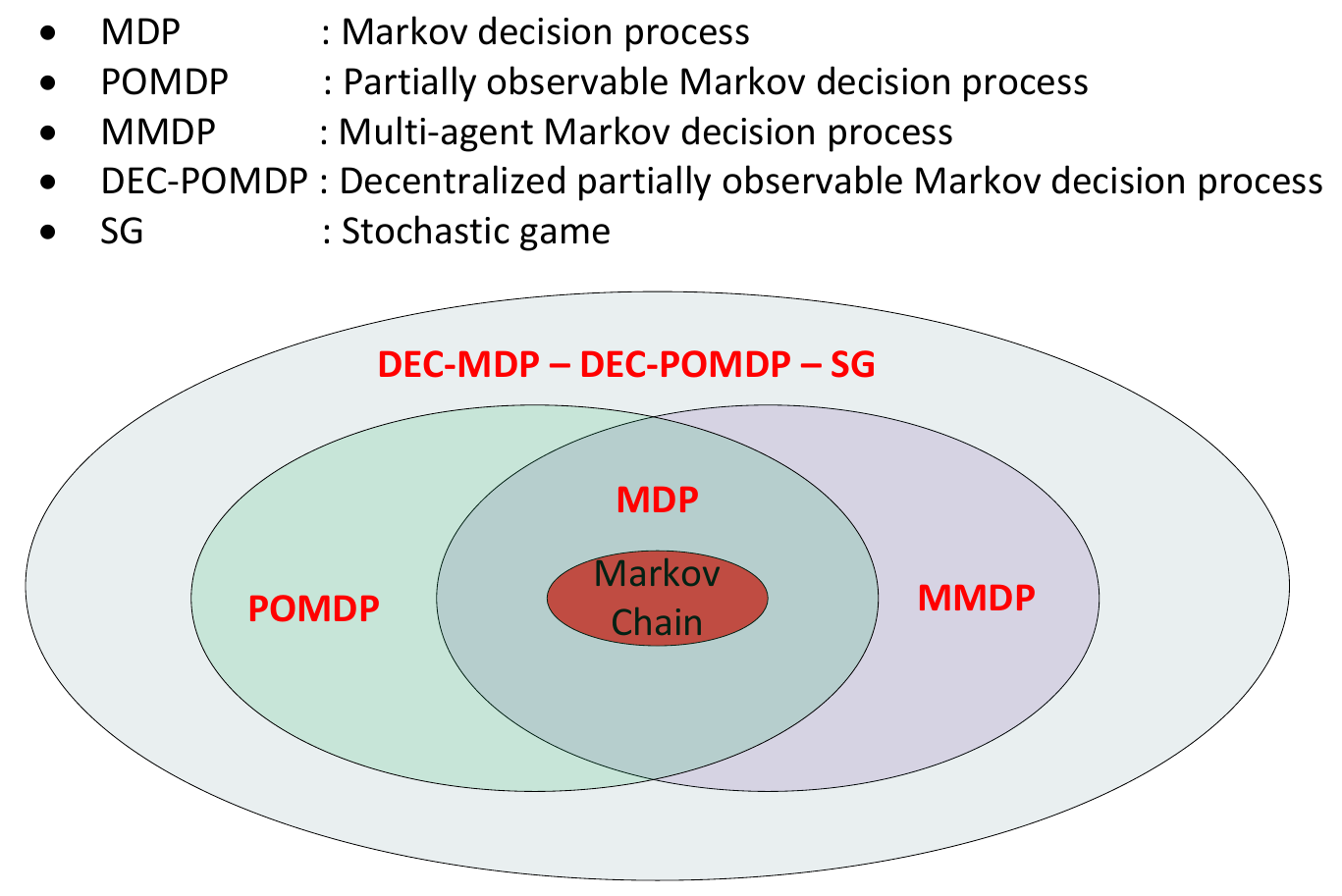}
\par\end{centering}
\caption{Extensions of Markov decision models.}
\label{fig:MDP_Models_comparison}
\end{figure}

Next we present some extensions of an MDP, the relation of which is shown in Figure~\ref{fig:MDP_Models_comparison}. 

\subsubsection{Partially Observable Markov Decision Processes (POMDPs)}
In classical MDPs, we assume that the system state is fully observable by an agent. However, in many WSNs, due to hardware limitations, environment dynamics, or external noise, the sensor nodes may not have full observability. Therefore, a POMDP~\cite{Monahan_1982_state_of_art} becomes an appropriate tool for such an incomplete information case. In POMDPs, the agent has to maintain the complete history of actions and observations in order to find an optimal policy, i.e., a mapping from histories to actions. However, instead of storing the entire history, the agent maintains a belief state that is the probability distribution over the states. The agent starts with an initial belief state $b_0$, based on which it takes an action and receives an observation. Based on the action and the received observation, the agent then updates a new belief state. Therefore, a POMDP can be transformed to an MDP with belief state~\cite{Chelsea_1991_A_survey, William_1991_A_survey}. Additionally, for a special case when the state space is continuous, parametric POMDPs~\cite{Brooks2007thesis} can be used. 
\subsubsection{Multi-Agent Markov Decision Processes (MMDPs)}
Unlike an MDP which is for a single agent, an MMDP allows multiple agents to cooperate to optimize a common objective~\cite{Boutilier_1999_sequential}. In MMDPs, at each decision time, the agents stay at certain states and they choose individual actions simultaneously. Each agent is assumed to have a full observation of the system state through some information exchange mechanism. Thus, if the joint action and state space of the agents can be seen as a set of basic actions and states, an MMDP can be formulated as a classical MDP. Thus, the solution methods for MDPs can be applied to solve MMDP. However, the state space and action space will drastically grow when the number of agents increases. Therefore, approximate solution methods are often used. 
\subsubsection{Decentralized Partially Observable Markov Decision Processes (DEC-POMDPs)}
Similar to MMDPs, DEC-POMDPs~\cite{Bernstein_2002_complexity} are for multiple cooperative agents. However, in MMDPs, each agent has a full observation to the system. By contrast, in DEC-POMDPs, each agent observes only part of the system state. In particular, the information that each agent obtains is local, making it difficult to solve DEC-POMDPs. Furthermore, in  DEC-POMDPs, because each agent makes a decision without any information about the action and state of other agents, finding the joint optimal policy becomes intractable. Therefore, the solution methods for a DEC-POMDP often utilize special features of the models or are based on approximation to circumvent the complexity issue~\cite{Oliehoek10PhD,Amato_2013_Decentralized}. Note that a decentralized Markov decision process (\emph{DEC-MDP}) is a special case of a DEC-POMDP that all agents share their observations and have a global system state. In WSNs, when the communication among sensors is costly or impossible, the DEC-POMDP is the best framework. 
\subsubsection{Stochastic Games (SGs)}

While MMDPs and DEC-POMDPs consider cooperative interaction among agents, stochastic games (or Markov games) model the case where agents are non-cooperative and aim to maximize their own payoff rationally~\cite{Shapley_1953_Stochastic}. In particular, agents know states of all others in the system. However, due to the different objective functions that lead to conflict among agents, finding an optimal strategy given the strategies of other agents is complex~\cite{Neyman_book_2003_Stochastic_games}. Note that the extension of stochastic games is known as a partial observable stochastic game~\cite{Hansen_2004_Dynamic} (\emph{POSG}) which has a fundamental difference in observation. Specifically, in POSGs, the agents know only local states. Therefore, similar to DEC-POMDPs, POSGs are difficult to solve due to incomplete information and decentralized decisions. 


It is proven that both finite time and infinite time horizon MDPs can be solved in complete polynomial time by dynamic programming~\cite{Christos_1987_The_complexity,Littman_1995_On_the}. However, extensions of MDPs may have different computation complexity. For example, for POMDPs, the agents have incomplete information and thus need to monitor and maintain a history of observations to infer the belief states. It is shown in~\cite{Goldsmith_1998_Complexity} that the complexity of POMDPs can vary in different circumstances and the worst case complexity is PSPACE-complete~\cite{Goldsmith_1998_Complexity, Christos_1987_The_complexity}. Since MMDPs can be converted to MDPs, its complexity in the worst case is P-complete. However, with multiple agents and partial observation (i.e., DEC-POMDP, DEC-POMDP, and POSG), the complexity is dramatically increased. It is shown in~\cite{Bernstein_2002_complexity} that even with just two independent agents, the complexity for both finite time horizon DEC-MDPs and DEC-POMDPs is NEXP-complete. Table~\ref{MDP_table_complexity} summarizes the worst case complexity. Note that partially observation problems are undecidable because infinite time horizon PODMPs are undecidable as shown in~\cite{Omid_2003_On_the}. 

\begin{table}[!]
\centering 
\caption{The Worst Case Complexity of Markov Models.}
\label{MDP_table_complexity}
    \begin{tabular}{| l | l | l |}
    \hline
	\textbf{\noun{Model}} & \textbf{\noun{Complexity}} \\ \hline \hline
	MDP & P-complete \tabularnewline\hline
	MMDP & P-complete \tabularnewline\hline
	POMDP (finite time horizon) & PSPACE-complete \tabularnewline\hline
	DEC-MDP (finite time horizon) & NEXP-complete \tabularnewline\hline 
	DEC-POMDP (finite time horizon) & NEXP-complete \tabularnewline\hline 
	POSG (finite time horizon) & NEXP-complete \tabularnewline\hline
   \end{tabular}
\end{table}

WSNs consist of tiny and resource-limited devices that cooperate to maintain the network topology and deliver the collected data to a data sink. However, the connecting links between nodes are not reliable and suffer from poor performance over time, e.g., fading effects. MDPs can model the time correlation in network structure and nodes. Therefore, many algorithms have been developed based on MDPs to address data exchange and topology maintenance issues. These methods are discussed in the next section.

\section{Data Exchange and Topology Formulation}\label{sec:Data-exchange}

A WSN may experience continual changes in its topology and transmission routes (e.g., new nodes can join the network, and existing nodes can encounter failures). This section reviews the applications of MDPs in data exchange and topology maintenance problems. Most surveyed works assume that the network consists of redundant sensors such that its operation can be performed by some alternative sensors. The use of MDPs in these applications can be summarized as follows: 
\begin{itemize} 
\item \emph{Data aggregation and routing}:  MDP models are used to obtain the most energy efficient sensor alternative for data exchange and gathering in cooperative multi-hop communications in WSNs. Different metrics can be included in the decision making such as transmission delay, energy consumption, and expected network congestion. 
\item \emph{Opportunistic transmission strategy}: Assuming sensors with adjustable transmission level, the MDP models adaptively select the minimum transmit power for the sensors to reach the destination. This adaptive transmission helps in reducing the energy consumption and the interference among nodes. 
\item \emph{Relay selection}: When the location and distance information is available at the source node, a relay selection decision can be optimized by using simple MDP-based techniques to reduce the energy consumption of the relay and source nodes. 
\end{itemize}

\subsection{Data Aggregation and Routing}

In WSNs, sensor nodes collect and deliver data to a data sink (e.g., a base station). Moreover, routing protocols are responsible for discovering the optimized routes in multi-hop transmissions~\cite{singh2010routing}. However, sensor nodes are resource-limited devices, and they can fail, e.g., because of hazardous environment. Accordingly, adaptive data aggregation and routing protocols are needed for WSNs. Different MDP models for such purposes are summarized in Table~\ref{tab:data_aggregation}. The column ``Decision'' specifies the decision making to be either distributed or centralized. Throughout the paper, we consider an algorithm to be distributed only if it does not require a centralized coordinator. Consequently, if the decision policy is computed using a central unit, and the policy is then disseminated to nodes, we still classify the algorithm as a centralized one. The columns ``States'', ``Actions'', and ``Rewards/costs'' describe the MDPs' components.

\begin{table*}
\centering{}\caption{\label{tab:data_aggregation}Data aggregation and routing techniques (SMDP = semi-MDP, E2E = end-to-end).}
\begin{tabular}{|c|>{\centering}p{1.1cm}|>{\centering}p{1cm}|>{\centering}p{1.2cm}|>{\centering}m{3cm}|>{\centering}m{3cm}|>{\centering}m{2.6cm}|}
\hline 
\textbf{\noun{Application context}} & \textbf{\noun{Article}} & \textbf{\noun{Type}} & \textbf{\noun{Decision}} & \textbf{\noun{States}} & \textbf{\noun{Actions }} & \textbf{\noun{Rewards/costs}}\tabularnewline
\hline 
\hline 
\multirow{2}{*}{Mobile networks} &~\cite{ye2009optimal}  & SMDP & Distributed & Arrivals of samples & Send, wait & Energy, E2E delay\tabularnewline
\cline{2-7} 
 &~\cite{fei2011efficient}  & MDP & Distributed & Transmission queue and distance & Select a moving direction & Data volume, the distance between the sensor and collector\tabularnewline
\hline 
Network query &~\cite{chobsri2009parametric} & POMDP & Centralized & Sensor's attribute values & Query (or do not query) a node & Query confidence\tabularnewline
\hline 
\multirow{4}{*}{The delay-energy tradeoff} &~\cite{lin2011autonomic} & MDP & Distributed & Channel's propagation gain, queue size & Select a transmit power & Received packets, E2E delay\tabularnewline
\cline{2-7} 
 &~\cite{hao2013energy} & MDP & Distributed & Forwarding candidates & Forward data, wait & Energy, E2E delay\tabularnewline
\cline{2-7} 
 &~\cite{guo2013opportunistic} & MDP & Distributed & Forwarding candidates & Select a transmit power and data forwarders & Energy, E2E delay\tabularnewline
\cline{2-7} 
 &~\cite{cheng2012adaptive} & MDP & Distributed & Available energy, mobile anchor's location & Select data forwarders & Energy variance, load balancing\tabularnewline
\hline 
\end{tabular}
\end{table*}

\subsubsection{Mobile Wireless Sensor Networks (MWSNs)}

Data collection in mobile sensor networks requires algorithms that capture the dynamics because of moving sensors, gateways, and data sinks. Moreover, distributed data aggregation can be even more challenging. Ye \textit{et al.}~\cite{ye2009optimal} addressed the problem of sensing a region of interest by exchanging data locally among sensors. This is referred to as a distributed data aggregation model, which also takes the tradeoff between energy consumption and data delivery latency into account. As data acquisition and exchange are stochastic processes in WSNs, the decision process is formulated as an SMDP with the expected total discounted reward. The model's states are characterized by the number of collected data samples by the node. This includes the samples forwarded by the neighbor nodes and the self-collected samples. The actions include (a) sending the queued data samples immediately, while stopping other operations, or (b) waiting until more data is collected. The waiting action can reduce a MAC control overhead when sending more data samples at once, achieving energy savings at the expense of increased data aggregation delay. Two real time solutions are provided, one is based on dynamic programming and the other is based on Q-learning. The interesting result is that the real time dynamic programming solution converges faster but consumes more computation resources than that of the Q-learning method.

In the similar context, Fei \textit{et al.}~\cite{fei2011efficient} formulated the data gathering problem in MWSNs using the MDP framework. Optimal movement paths are defined for mobile nodes (i.e., data sinks) to collect sensor readings. The states are the locations of the mobile nodes. The region of interest is divided into a grid, and each node decides to move to one of the nine possible segments. The reward function reflects the energy consumption of the node and the number of collected readings. Numerical results show that the proposed scheme outperforms conventional methods, such as the traveling salesman-based solutions, in terms of connectivity error and average sensing delay.

Even though MWSNs have some useful properties over static networks, some of their drawbacks must be considered. Basically, these MWSN algorithms are hard to implement and maintain in real world scenarios, and distributed MDP algorithms converge after a long-lived exploration phase which could be costly for resource-limited devices.

\subsubsection{Network Query}

Data query in WSNs serves to disseminate a command (i.e., a query) from a base station to the intended sensor nodes to retrieve their readings. Chobsri \textit{et al.}~\cite{chobsri2009parametric} proposed a probabilistic scheme to select the set of sensor nodes that should answer to the user query. For example, a query may request for the data attributes, e.g., temperature, to be within an intended confidence range. The problem is formulated as a parametric POMDP with average long-term rewards as the optimization metric. The action of the node is whether to answer the query or not. The states are formulated as a vector that includes the data attribute from each sensor. Since the sensors are error prone, the base station maintains the collected readings as beliefs (i.e., not the actual states). The data acquisition problem is solved using the value iteration method to achieve near-optimal selection policy. 

\subsubsection{Delay-Energy Tradeoff}

In~\cite{lin2011autonomic}, Lin \textit{et al.} suggested a distributed algorithm for delay-sensitive WSNs under dynamic environments. The environment is considered stochastic in terms of the traffic pattern and wireless channel condition. A transceiver unit of a sensor node controls its transmission strategies, e.g., transmit power levels, to maximize the node's reward. In each node, wireless channel's propagation gain and queue size are used to define the MDP's states. The actions consider the joint power consumption (i.e., transmission power level) and the next forwarder selection decisions. Additionally, the messages are prioritized for transmission using the earliest deadline first criterion. Similarly, Hao \textit{et al.}~\cite{hao2013energy} studied the energy consumption and delay tradeoff in WSNs using the MDP framework. The nodes select their actions of ``immediate data forwarding'' or ``wait to collect more data samples'' based on local network states (i.e., the number of relay candidates). The local network states include the node's own duty cycle (i.e., activation mode) and its neighbor's duty cycle modes. Furthermore, the duty cycle of the nodes is managed using simple beacon messages exchanged locally among the nodes to inform each other about their wake up (i.e., active) mode. The numerical results show that the adaptive routing protocol enhances successful packet delivery ratios under end-to-end delay constraints.

\begin{figure}
\begin{centering}
\includegraphics[width=0.75\columnwidth,trim=1cm 1cm 1cm 0.5cm]{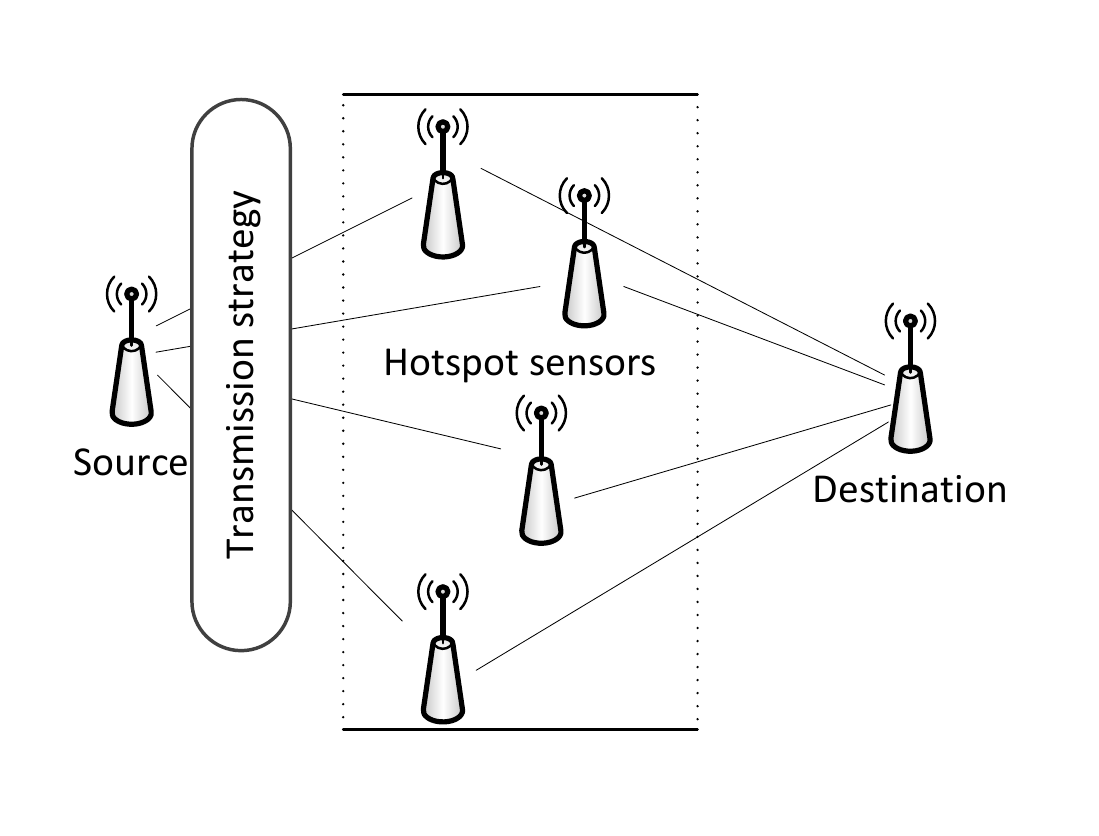}
\par\end{centering}

\caption{\label{fig:cheng2012adaptive}Data transmission strategy to decide sending data over one of the available paths as considered in~\cite{cheng2012adaptive}.}
\end{figure}

Guo \textit{et al.}~\cite{guo2013opportunistic} used MDPs to develop an opportunistic routing protocol for WSNs with controlled transmit power level. The preferred power source is selected by finding the optimal policy of the MDP's configuration. Again, each potential next forwarding node is considered as a state in the MDP, and source and destination nodes are considered as the initial and goal states, respectively. Compared with conventional routing protocols, the proposed scheme shortens the end-to-end delay and consumes less energy as a result of the opportunistic routing decisions. Furthermore, in~\cite{cheng2012adaptive}, Cheng and Chang suggested a solution to manage node selection in event-detection applications using mobile nodes equipped with directional antenna and global positioning system (GPS). Basically, the directional antenna and GPS technologies are used to divide the network into operational zones. The solution aims at balancing the energy consumption of the next forwarding nodes surrounding the sink, i.e., the energy of the hotspot nodes (Figure~\ref{fig:cheng2012adaptive}). The fully observable states are based on the energy level and positions of the nodes within the hotspot. The discounted reward is formulated to find an optimal action for selecting the data forwarding node at each time instant. In this solution, transition probabilities are not needed, as reinforcement learning is used to solve the formulated MDP model.

Overall speaking, fully observable MDPs have been successfully applied to find a balanced delay-energy tradeoff as shown in~\cite{lin2011autonomic,hao2013energy,guo2013opportunistic,cheng2012adaptive}. However, only limited comparison to other state-of-the-art algorithms is provided in these papers, which restricts the result interpretation and performance gain evaluation.

\subsection{Opportunistic Transmission Strategy and Neighbor Discovery}

Opportunistic transmission with neighbor discovery is an essential technique in large scale WSNs to achieve the minimum transmit power that is needed to maintain the network connectivity among neighboring nodes and exchange discovery messages. In addition to minimizing the energy consumption, opportunistic transmission is also important to minimize data collision among concurrent data transmission. The transmit power is also defined to meet the signal-to-noise ratio (SNR) requirements. Moreover, channel state information (CSI) is a widely used indicator for the channel property and signal propagation through channels. A summary of surveyed MDP-based transmission methods is given in Table~\ref{tab:transmission_strategies}.

\begin{table*}
\caption{\label{tab:transmission_strategies}Transmission strategies and neighbor discovery methods (SG = stochastic game, CSI = channel state information, SNR = signal to noise ratio, E2E = end-to-end).}

\begin{centering}
\begin{tabular}{|c|>{\centering}m{1.1cm}|>{\centering}p{1cm}|>{\centering}p{1.2cm}|>{\centering}m{3cm}|>{\centering}m{3cm}|>{\centering}m{2.6cm}|}
\hline 
\textbf{\noun{Application context}} & \textbf{\noun{Article}} & \textbf{\noun{Type}} & \textbf{\noun{Decision}} & \textbf{\noun{States}} & \textbf{\noun{Actions}} & \textbf{\noun{Rewards/costs}}\tabularnewline
\hline 
\hline 
\multirow{4}{*}{Transmission policies} &~\cite{pandana2005near} & MDP & Distributed & Buffer occupancy, SIR, data rate & Select modulation level and transmit power & Received packets, energy\tabularnewline
\cline{2-7} 
 &~\cite{krishnamurthy2005game} & SG & Distributed & CSI & Transmit, wait & Delay, energy, successful transmission\tabularnewline
\cline{2-7} 
 &~\cite{madan2006energy} & MDP & Distributed & Cones with at least one neighbor node & Select a transmit power & Energy, neighbor discovery\tabularnewline
\cline{2-7} 
 &~\cite{stabellini2008energy} & MDP & Distributed & Cones with at least one neighbor node & Select a transmit power & Collision, energy, neighbor discovery\tabularnewline
\hline 
\multirow{4}{*}{Transmission scheduling} &~\cite{boloni2008should} & MDP & Distributed & Buffer occupancy, distance to mobile sink & Transmit using static nodes or wait for a mobile sink & E2E delay, energy\tabularnewline
\cline{2-7} 
 &~\cite{van2010energy} & MDP & Distributed & SNR, channel's Doppler frequency & Transmit, wait & Energy, collision\tabularnewline
\cline{2-7} 
 &~\cite{xiong2010cross} & MDP & Distributed & CSI, buffer occupancy, transmission success probability & Select cross layer transmission protocols & Energy\tabularnewline
\cline{2-7} 
 &~\cite{xiong2012channel} & MDP & Distributed & CSI, transmission success probability & Transmit, wait & Energy\tabularnewline
\hline 
\multirow{3}{*}{Transmit power} &~\cite{udenze2008partially} & POMDP & Distributed & Channel states observed by transmission outcome  & Transmit data, wait, probe the channel & Interference, energy\tabularnewline
\cline{2-7} 
 &~\cite{kobbane2012dynamic} & MDP & Centralized & Residual energy & Select a transmit power & Energy, throughput\tabularnewline
\cline{2-7} 
 &~\cite{gatsis2013optimal} & MDP & Centralized & Fading channel coefficient, reception error  & Select a transmit power & Reception probability, energy\tabularnewline
\hline 
\end{tabular}
\par\end{centering}

\end{table*}

\subsubsection{Distributed Transmission Policies}

Pandana and Liu~\cite{pandana2005near} presented an adaptive and distributed transmission policy for WSNs. The policy examines the signal-to-interference ratio (SIR), data generation rate at source sensors, and the data buffer capacity. The solution is also suitable for data exchange among multiple nodes. Reinforcement learning is used to solve the MDP formulation, which results in near-optimal estimation of transmit power and modulation level. The reward function presents the total number of successful transmissions over total energy consumption. Therefore, an optimized transmission policy requires using a suitable power level and data buffer without overflow. The suggested scheme is compared with a simple policy which selects a transmission decision (a modulation level and transmit power) to match the predefined SIR requirement. The experiment shows that the proposed scheme achieves twice the throughput of the simple policy.

Krishnamurthy \textit{et al.}~\cite{krishnamurthy2005game} considered the slotted ALOHA protocol in WSNs with the aim to maximize the network throughput using stochastic games (SGs). Specifically, each node tries to minimize its transmission collision with other non-cooperative nodes and by exploiting only the CSI. The players are the nodes with two possible transmission actions of waiting and immediate channel access. Then, the intended policy probabilistically maps CSI conditions, i.e., states, to the transmission action. Using a distributed threshold policy, the nodes achieve a Nash equilibrium solution, where the game formulation assumes finite numbers of players and actions. The experiments reveal that the threshold value is proportional to the number of nodes, and therefore each node is less probable to access the channel when the network size increases.

In light of previous studies, Madan and Lall~\cite{madan2006energy} considered the problem of neighbor discovery of randomly deployed nodes in WSNs. The node deployment over an area is assumed to follow the Poisson distribution. An MDP model is used to solve the neighbor discovery problem to minimize energy consumption. The plane area surrounding each node is divided into cones (e.g., 3 cones) and the neighbor discovery algorithm must ensure that there is at least one connected neighbor node in each cone. To minimize the computational complexity, the MDP policy is solved offline, using linear or dynamic programming methods. Then, the policy is stored on the nodes for an online operation. In the MDP formulation, states are the number of connected cones and the discrete levels of transmit power in previous interval. The nodes manage the minimum required transmit power (i.e., the MDP actions) to discover the neighbor nodes. In~\cite{stabellini2008energy}, Stabellini \textit{et al.} extended~\cite{madan2006energy} and proposed an MDP-based neighbor discovery algorithm for WSNs that is solved using dynamic programming. Unlike~\cite{madan2006energy}, the energy consumption of the node in listening mode is considered. The model proposed in~\cite{madan2006energy} considers the average energy consumption which monotonically decreases as the node density increases. This does not take into account the contention windows and collisions in dense networks. This modeling limitation is solved in~\cite{stabellini2008energy} by considering any additional transmissions that result from undiscovered neighbors.

A primary limitation of the algorithms presented in~\cite{pandana2005near,krishnamurthy2005game,madan2006energy,stabellini2008energy} is their discrete models (i.e., transmission decisions are only made at discrete time intervals). This means that a node must stay in a state for some time before moving to the next state which hinders the use of the algorithm in critically time-sensitive applications. An interesting research direction would be in using continuous time models and SMDPs for proposing distributed transmission policies. 

\subsubsection{Packet Transmission Scheduling}

In~\cite{boloni2008should}, Bölöni and Turgut introduced two scheduling mechanisms for packet transmission in energy-constrained WSNs with mobile sinks. Occasionally, a static node may not be able to aggregate its data using a nearby mobile sink, and can use only the more expensive multi-hop retransmission method for data aggregation. Thus, the scheduling mechanism decides if the node should wait for a mobile sink to move and come into proximity, or immediately transmit data through the other static nodes. The first mechanism uses a regular MDP method, and the second one introduces historical data to sequential state formulation. Thus, the former method (i.e., without using historical data) outperforms the latter, despite not having precise knowledge of the sink node mobility pattern. Likewise, Van Phan \textit{et al.}~\cite{van2010energy} addressed the transmission strategy optimization, while minimizing the energy consumed for unsuccessful transmission. The SNR is used to predict the channel states (i.e., good or bad), by using a simple threshold mechanism. The transition probabilities can be calculated using the channel Doppler frequency, the frame transmission time, and the probability of symbol error. A transmission is performed only when the channel is good, which can increase the probability of success. Simulations using the Network Simulator (NS-2) and fading channels with 20 states show the energy efficiency of the introduced transmission policy.

Xiong \textit{et al.}~\cite{xiong2010cross} proposed an MDP-based redundant transmission scheme for time-varying channels. The data redundancy can achieve better performance and lower energy consumption than that of conventional retransmission schemes especially in harsh environments. In this case, each node estimates the energy saving of its contribution on forwarding data packets. The algorithm selects the optimized cross-layer protocols to transmit the data at the current condition, e.g., combining the forward error correction (FEC) and automatic repeat request (ARQ) protocols. The CSI, extracted at the physical layer, is considered as states. This cross-layer solution formulates the cost as a function of energy consumption in an infinite horizon time domain. Again and unlike the data redundancy method used in~\cite{xiong2010cross}, Xiong \textit{et al.}~\cite{xiong2012channel} tackled the design of optimal transmission strategy, while data retransmission is performed for undelivered packets.

\subsubsection{Wireless Transmit Power}

\begin{figure}
\begin{centering}
\includegraphics[width=0.65\columnwidth,trim=1cm 1cm 1cm 0.5cm]{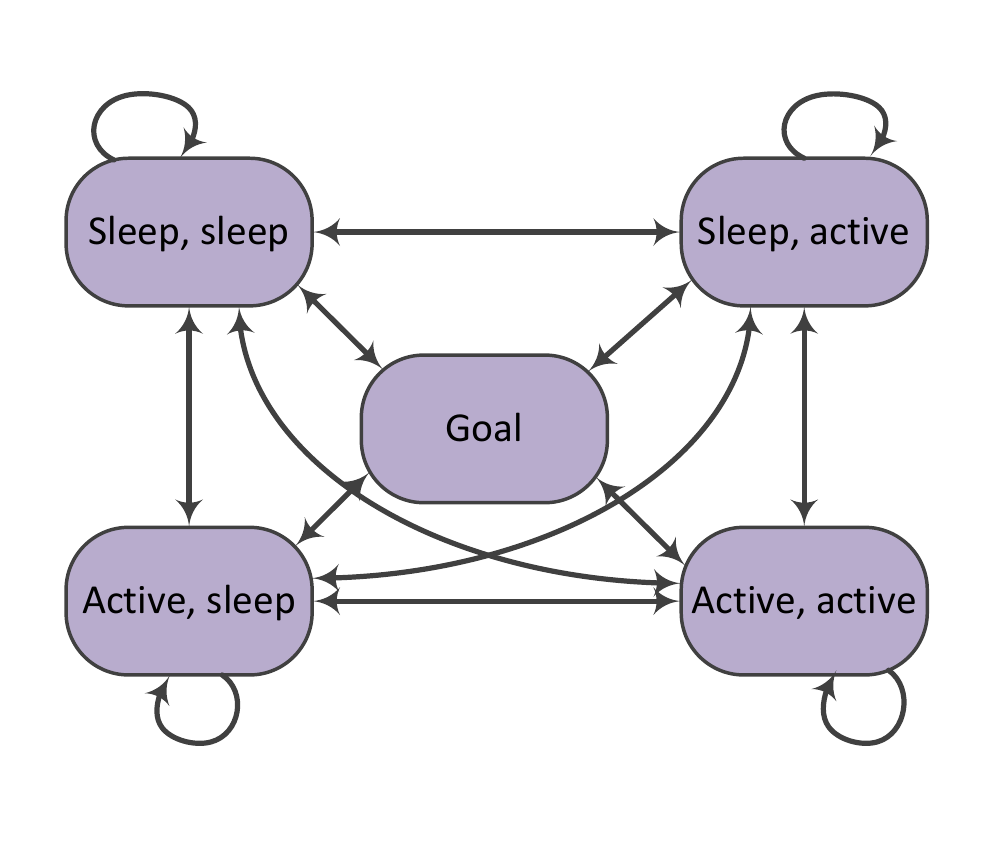}
\par\end{centering}

\caption{\label{fig:udenze2008partially}State transition of two nodes under the scheme proposed in~\cite{udenze2008partially}.}
\end{figure}

Udenze and McDonald-Maier~\cite{udenze2008partially} presented a POMDP-based method to manage transmit power and transmission duration in WSNs, while adapting with the system dynamics, e.g., unknown interference model. The partial information about interfering nodes is used to define the problem observations and beliefs. For example, a successful transmission indicates an idle channel state. Each node has partial information about the environment as a hidden terminal problem may exist. Figure~\ref{fig:udenze2008partially} shows an example of the allowable state transition of two nodes. The node decides to transmit data at a specific energy level, continue waiting, or send a probing message to test the channel status. Thus, each node can utilize channel information to increase its transmission probability during the channel idle state.

\begin{figure}
\begin{centering}
\includegraphics[width=0.75\columnwidth,trim=1cm 1cm 1cm 0.5cm]{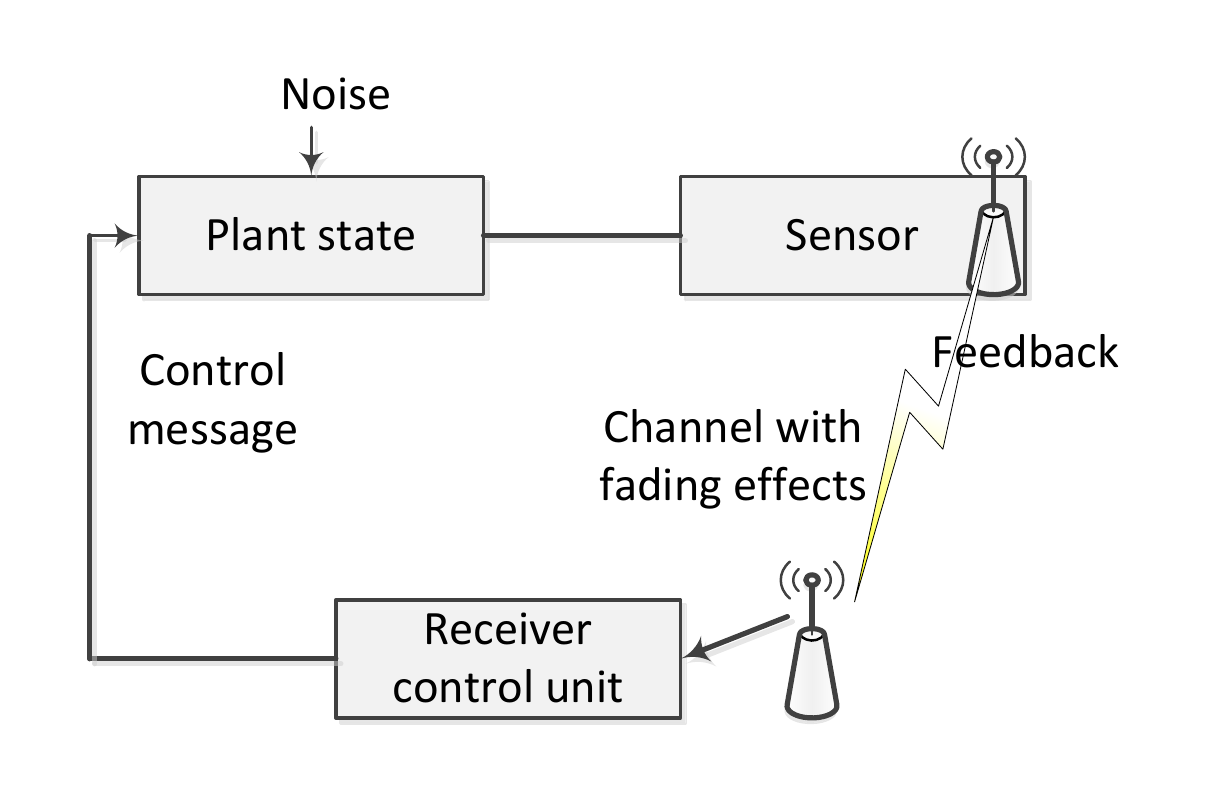}
\par\end{centering}

\caption{\label{fig:gatsis2013optimal}System architecture of the closed loop control system tackled in~\cite{gatsis2013optimal}.}
\end{figure}

Kobbane \textit{et al.}~\cite{kobbane2012dynamic} built an energy configuration model using an MDP. This centralized scheme is to manage the node transmission behavior to maximize the network's lifetime, while ensuring the network connectivity and operations. The backend (e.g., a base station), which runs the centralized scheme, is assumed to have complete information about the environment including the nodes' battery levels and connecting channel states. As a centralized method, no local information exchange is required among the sensors, as the nodes receive the decision policy from the backend. Based on the simulation with 64 states, the interesting result is that the transmit power policy takes constant values during the first 40 time slots of the simulation, and subsequently the transmit power increases as the state value increases. A more specialized framework was proposed by Gatsis \textit{et al.}~\cite{gatsis2013optimal}  to address the transmit power control problem in WSNs used for dynamic control systems. Intuitively, the more the gathered information from sensors, the more precise the decision can be made at the control unit. However, this increases energy consumption at the nodes. In the infinite time horizon formulation, the MDP considers the reception (decoding) error and the channel fading condition which are determined by a feedback controller as shown in Figure \ref{fig:gatsis2013optimal}. Thereafter, suitable transmit power can be selected to achieve a functional control system operation at a minimum operating cost.

\subsection{Relay Selection and Cooperative Communications}

A source sensor node has to select relay node(s) to forward data to an intended sink. This is based on the maximum transmission range of the source node, and available energy of the source and relay nodes. Relay placement is usually designed to keep the network connected using the minimum set of relay nodes~\cite{cheng2008relay}. The source node may use direct transmission mode, if applicable, to save the overall network energy when it cannot find a suitable relay node. Thus, the relay selection problem must evaluate the energy consumption of relay paths and direct link decisions. MDPs are employed in relay selection and cooperative networking as summarized in Table~\ref{tab:relay_selection}.

\begin{table*}
\caption{\label{tab:relay_selection}Relay selection and cooperative communications (E2E = end-to-end).}

\centering{}%
\begin{tabular}{|c|>{\centering}m{1.1cm}|>{\centering}m{1cm}|>{\centering}m{1.2cm}|>{\centering}m{2.8cm}|>{\centering}m{2.8cm}|>{\centering}m{2.5cm}|}
\hline 
\textbf{\noun{Application context}} & \textbf{\noun{Article}} & \textbf{\noun{Type}} & \textbf{\noun{Decision}} & \textbf{\noun{States}} & \textbf{\noun{Actions}} & \textbf{\noun{Rewards/costs}}\tabularnewline
\hline 
\hline 
\multirow{2}{*}{Relays with energy harvesting} &~\cite{li2011relay} & POMDP, MDP & Distributed & Energy budget, event occurrence & Transmit directly or use a relay & Data priority, energy, coverage\tabularnewline
\cline{2-7} 
 &~\cite{li2012analytical} & POMDP & Distributed & Available energy, event occurrence, recharging state & Transmit directly or use a relay & Accuracy, energy\tabularnewline
\hline 
\multirow{4}{*}{Relay activation and scheduling} &~\cite{koulali2012optimal} & MDP & Distributed & Duty cycle of nodes within transmission range & Transmit or wait & E2E delay, energy\tabularnewline
\cline{2-7} 
 &~\cite{naveen2012relay} & MDP & Distributed & Relay set & Transmit, wait, or probe for other relays & E2E delay, energy\tabularnewline
\cline{2-7} 
 &~\cite{naveen2013relay} & POMDP & Distributed & Relay set observed by wake-up instants & Transmit or wait & Hop count, E2E delay, energy\tabularnewline
\cline{2-7} 
 &~\cite{sinha2014optimal} & MDP & Centralized & Relative deployment location & Place (or do not place) a relay & Hop count\tabularnewline
\hline 
\end{tabular}
\end{table*}

\subsubsection{Relay Selection with Energy Harvesting}

In~\cite{li2011relay}, Li \textit{et al.} addressed the relay selection problem in energy harvesting sensor nodes. The problem is formulated as a POMDP and relaxed to an MDP for obtaining the solution. The states of source and relay nodes are characterized by energy budgets and any event occurrence. Naturally, the battery budget varies because of energy consumption and recharging processes. The source node fully observes its own state but has partial information on the other relay nodes. Every node decides if it should participate in the current transmission to maximize the average reward. The available actions of the communicating devices (a source and a relay node) are ``idle, idle'', ``direct, idle'', ``relay, relay'', ``direct, self-traffic'', and ``idle, self-traffic''. Again, Li \textit{et al.}~\cite{li2012analytical} reused the POMDP formulation previously proposed in~\cite{li2011relay}, however, with the intention of providing a practical and near-optimal solution. In particular, they consider the tradeoff between solution optimality and computational complexity. The state space comprises the available energy, event occurrence, and recharging state (on or off). The actions are similar to those in~\cite{li2011relay}. Relay selection is only explored once the source node's energy budget is below a defined threshold. This naive method, i.e., the threshold mechanism, is shown to provide near-optimal solution with low computational complexity. Running a simulation test case of 5 million time units shows that the threshold based scheme consumes only half of the energy of the optimal policy solution while achieving near-optimal packet delivery ratio.

The main limitation of~\cite{li2011relay, li2011relay} is the low performance when operating in harsh environments, e.g., because of rapidly changing channel interference. In such cases, the relay selection policy has to be reconstructed to fit the new conditions which will be a resource demanding task.

\subsubsection{Relay Activation and Scheduling}

Koulali \textit{et al.}~\cite{koulali2012optimal} proposed an MDP-based scheduling mechanism in WSNs by modeling sensors' wake up patterns. A sensor wakes up either to sense a sporadic event or to relay other nodes' messages. A relaying node can transmit the data to the already active next hop node, or it waits for the activation of other nodes nearer to the sink. Therefore, the tradeoff between data gathering delay and expected network energy consumption is considered. A node can be either in the active or sleep mode.

Naveen and Kumar~\cite{naveen2012relay} extended previous studies that tackled relay selection in WSNs using an MDP. In particular, in addition to being able to select among the explored relay nodes, a transmitting node can decide to continue probing to search for farther relay options. During the probing, the node determines the reward to be distributed to the reachable relays. The states are the best historical reward and the rewards of un-probed relays at previous stages. Then, the Bellman equation is used to solve the MDP formulation. Subsequently, Naveen and Kumar~\cite{naveen2013relay} discussed geographical data transmission in event-monitoring WSNs. As long as the nodes' duty cycles are asynchronous, the nodes need to control the sleep time, i.e., wait or wake up for transmission, to match that of their relay neighbors. The waiting time of the nodes and the progress of data forwarding toward the sink are employed in the state of the POMDP. The partial observability in the system is introduced as the number of relays is assumed to be unknown as no beacon message is exchanged between neighboring nodes. 

Sinha \textit{et al.}~\cite{sinha2014optimal} discussed the online and random construction of relay paths from source node to destinations. The solution explores the placement of relays to minimize the weighted cost including the hop count and transmission costs. This MDP model is independent of location, and it considers only the previous relay placement to predict the optimal deployment of the next relay. The model is useful in densely covered regions, e.g., forests. However, the online placement of relays can be used only in very low rate networks, e.g., one reading over a few seconds. The extraction of the optimal policy requires a finite number of iterations which makes this solution suitable for WSNs. However, the conventional placement methods that are based on a distance threshold can achieve near-optimal results when the threshold value is carefully selected.

When dealing with relay activation and scheduling, the best suited MDP variant is the POMDP model because of the low communication overhead as shown in ~\cite{naveen2013relay}. However, other algorithms (e.g., ~\cite{koulali2012optimal, naveen2012relay,sinha2014optimal}) assume the full information availability about the neighboring nodes when making decisions.

In summary, there are two important remarks about the reviewed algorithms in this section for data exchange and topology formation. Firstly, the fully observable MDP model with complete information about neighbor nodes and relays has been favored in most reviewed papers. This is due to the low computational burden of the fully observable model. However, this is at the cost of increased transmission overhead as exchanging beacon messages is required. Secondly, the reviewed papers have clearly shown the efficiency of the MDP models in the problems related to data exchange and topology formulation. However, most of these papers do not include long-running experiments using real world testbeds and deployments to assess their viability and system-wide performance under changing conditions. The next section discusses the use of MDPs in resource and power optimization algorithms.

\section{Resource and Power Optimization}\label{sec:Resource-and-power}

A major issue of WSN design is the resource usage at the node level. Resources include available energy, wireless bandwidth, and computational power. Clearly, most of the surveyed papers in this section focus on resource-limited nodes and long network lifetime. In particular, the related work uses MDP for the following studies: 
\begin{itemize} 
\item \emph{Energy control}: For the energy charging of sensors, an MDP is used to decide on the optimal time and order of sensor charging. These energy recharging methods consider each node's battery charging/discharging time and available energy. Moreover, some of them deal with the stochastic nature of energy harvesting in WSNs. Therefore, the energy harvesting nodes are scheduled to perform tasks that fit their expected energy harvesting. 
\item \emph{Dynamic optimization}: A sensor node should optimize its operation at all protocol stacks, e.g., data link and physical layers. This is to match the physical conditions of the environment, e.g., weather. Therefore, unlike static configurations, the operations are performed at minimized resource consumption, while providing service in harsh conditions. Moreover, MDP-based network maintenance algorithms were developed. These maintenance models generate a low cost maintenance procedure, while assuring network functionality over time.  
\item \emph{Duty cycling and channel access scheduling}: Sensor nodes consume less energy when they switch to a sleep mode. The MDP-based methods predict the optimal wake up and sleep patterns of the sensors. During duty cycle management, each node evaluates the activation of surrounding nodes to exchange data and to minimize the interference with other nodes. 
\end{itemize}

\subsection{Energy Recharging, Harvesting, and Control}

The literature is rich with MDP-based energy control as summarized in Table~\ref{tab:energy_control}. These solutions consider the energy recharging and harvesting in WSNs as follows.

\begin{table*}
\caption{\label{tab:energy_control}Comparison among energy control and harvesting techniques (CMDP = constrained Markov decision process).}

\centering{}%
\begin{tabular}{|c|>{\centering}m{1.1cm}|>{\centering}m{1cm}|>{\centering}m{1.2cm}|>{\centering}m{3cm}|>{\centering}m{3cm}|>{\centering}m{2.6cm}|}
\hline 
\textbf{\noun{Application context}} & \textbf{\noun{Article}} & \textbf{\noun{Type}} & \textbf{\noun{Decision}} & \textbf{\noun{States}} & \textbf{\noun{Actions}} & \textbf{\noun{Rewards/costs}}\tabularnewline
\hline 
\hline 
\multirow{2}{*}{Recharging management} &~\cite{misra2010markov} & MDP & Centralized & Quantized available energy & Recharge the battery (sleep) or continue operations & Recharging delay, network disruption\tabularnewline
\cline{2-7} 
 &~\cite{osais2010optimal} & MDP & Centralized & Available energy, sensor temperature & Recharge, sleep, or sample & Number of collected samples\tabularnewline
\hline 
\multirow{11}{*}{Energy harvesting} &~\cite{park2012optimal} & POMDP & Distributed & Spectrum state, available energy & Access the spectrum or wait & Successful packet delivery\tabularnewline
\cline{2-7} 
 &~\cite{kashef2012optimal} & MDP & Distributed & Spectrum state, available energy & Access the spectrum, or wait & Successful packet delivery\tabularnewline
\cline{2-7} 
 &~\cite{iannello2012optimality} & POMDP & Centralized & Available energy, transmission outcome & Schedule the spectrum access & Successful packet delivery\tabularnewline
\cline{2-7} 
 &~\cite{mohamed2013adaptive} & MDP & Centralized & Available energy, expected energy harvesting, event occurrence, buffer
occupancy & Set the compression error & Compression accuracy, energy\tabularnewline
\cline{2-7} 
 &~\cite{michelusi2013transmission} & MDP & Centralized & Available energy, harvesting state, data importance & Transmit or discard packets & Delivery of important packets\tabularnewline
\cline{2-7} 
 &~\cite{aprem2013transmit} & POMDP & Centralized & Partial CSI, available energy, data packets & Select a transmit power & Successful packet delivery\tabularnewline
\cline{2-7} 
 &~\cite{niyato2013delay} & CMDP & Distributed & E2E delay, available energy, mobile location & Transmit or continue waiting & Deadline violation\tabularnewline
\cline{2-7} 
 &~\cite{murtaza2013optimal} & MDP & Distributed & Weather condition, available energy & Transmit or continue waiting & Transmission rate, charging rate\tabularnewline
\cline{2-7} 
 &~\cite{mao2014joint} & MDP & Distributed & Available energy, buffer occupancy, harvesting state, channel state & Allocate energy for transmission and sensing & Successful packet delivery\tabularnewline
\cline{2-7} 
 &~\cite{nourian2014optimal} & MDP & Centralized & Available energy, channel gain & Select a transmit power & Packet dropping\tabularnewline
\cline{2-7} 
 &~\cite{ren2014dynamic} & MDP, POMDP & Distributed & Event occurrence, available energy,

node's activation history & Activate or sleep & Energy, event detection\tabularnewline
\hline 
\end{tabular}
\end{table*}

\subsubsection{Recharging Management}

In WSNs, a sensor node may have to operate without battery replacement. Moreover, the nodes drain their energy unequally because of different operation characteristics. For example, the nodes near the data sink drain energy faster as they need to relay other nodes' data. The battery charging of the node must be performed to fit the node's energy consumption and traffic load conditions. Accordingly, an MDP is used to select the order and the time instant of node charging. Note that the node charging can be based on wired and wireless energy transfer.

Misra \textit{et al.}~\cite{misra2010markov} used an MDP to model the energy recharging process in WSNs. Naturally, since the available energy levels affect the recharging delay, the recharging process of nodes must be designed to account for the difference in available energy at different nodes. The available energy of different nodes differs because of different transmission history and different battery technologies used in the nodes. Thus, the recharging process of the nodes is also not a uniform task, and some nodes need longer charging time than others. Therefore, the proposed solution is intended to minimize the recharge delay and maximize the total number of recharged nodes. The battery budget is quantized into a few states, e.g., $\left\{ [0\%-20\%],\ldots,[80\%-100\%]\right\} $. At each decision interval, the node decides either to perform energy charging under sleep mode, or to continue its active mode (recharging cannot be done under the active mode). 

In~\cite{osais2010optimal}, Osais \textit{et al.} discussed the problem of managing the recharging procedure of nodes attached to human and animal bodies. Therefore, it is important to take the temperature produced from the inductive charging of batteries into account, as a high temperature harms the body. Under the maximum threshold of acceptable temperature, the proposed solution produces an MDP policy that maximizes the number of readings collected by any node, and therefore enhances the body monitoring. The state of the  node is characterized by its current temperature and energy level. At each interval, an action is selected from three feasible options: (i) recharge the battery, (ii) switch to sleep mode, or (iii) collect data sample. A heuristic policy is proposed to minimize the computational complexity of the optimal policy. In short, the heuristic policy selects actions based on the current biosensor's temperature and energy level. For example, the sample action is chosen at low temperature values and high energy levels, while the recharge action is performed at very low energy levels. The heuristic policy is compared with a greedy policy. The greedy policy selects an action based on a fixed-priority order: sample, recharge and sleep. The simulation shows that the heuristic policy's performance is close to that of an optimal policy, and it reduces the sample generation time by 75\% when compared with the greedy approach.

An interesting extension of~\cite{misra2010markov,osais2010optimal} is to consider event occurrence and data priority which enables the delivery of important packets even at low available energies. Another appealing future research direction is to implement distributed algorithms using partial information models to minimize the transmission overhead.

\subsubsection{Energy Harvesting}

\begin{figure}
\begin{centering}
\includegraphics[width=0.85\columnwidth,trim=1cm 1cm 1cm 0.5cm]{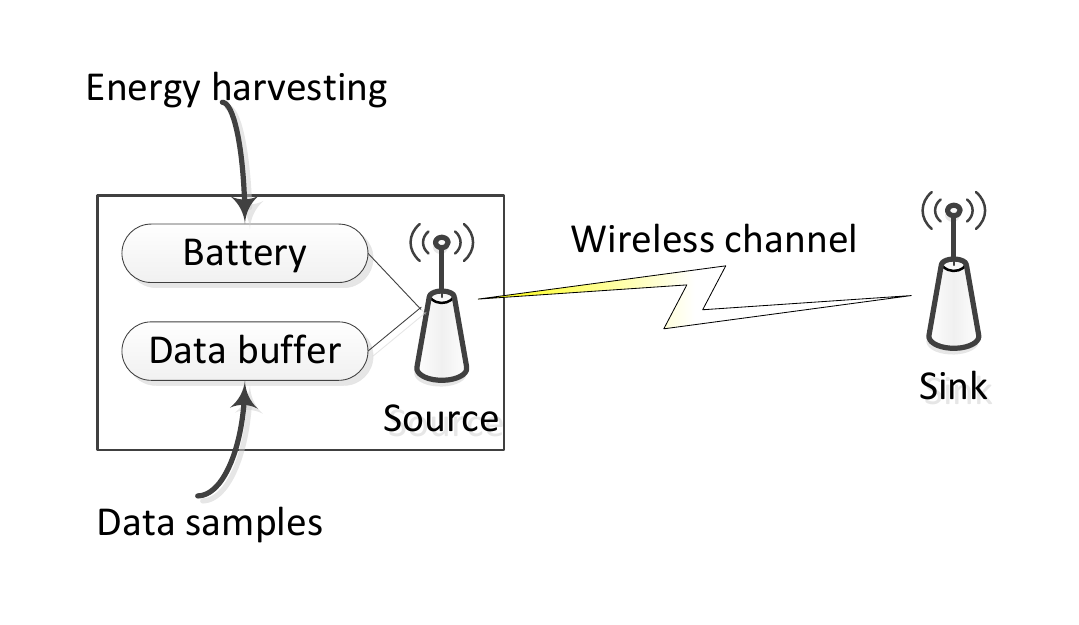}
\par\end{centering}

\caption{\label{fig:energy_harvesting}System model of energy harvesting methods.}
\end{figure}

Battery charging can be complex and inconvenient in many cases. Therefore, a more viable choice is to harvest energy from the environment, e.g., thermal and radiant energy, for a sensor node's battery~\cite{sudevalayam2011energy}. Although the natural energy is free and infinite, it is random and sporadic. Therefore, many studies explored the prediction of energy harvesting in WSNs. The majority of research efforts in the literature examine the dynamics of available energy and buffer size as shown in Figure \ref{fig:energy_harvesting} to optimize node's operations. Thereby, a balanced tradeoff between the energy consumption and harvesting is achieved. We refer the readers to~\cite{kausar2014energizing,valera2014survey} for more insight on energy harvesting in WSNs and its challenges. Instead, here we focus on the applications of MDPs for energy harvesting in WSNs.

\begin{figure}
\begin{centering}
\includegraphics[width=1\columnwidth,trim=1cm 1cm 1cm 0.5cm]{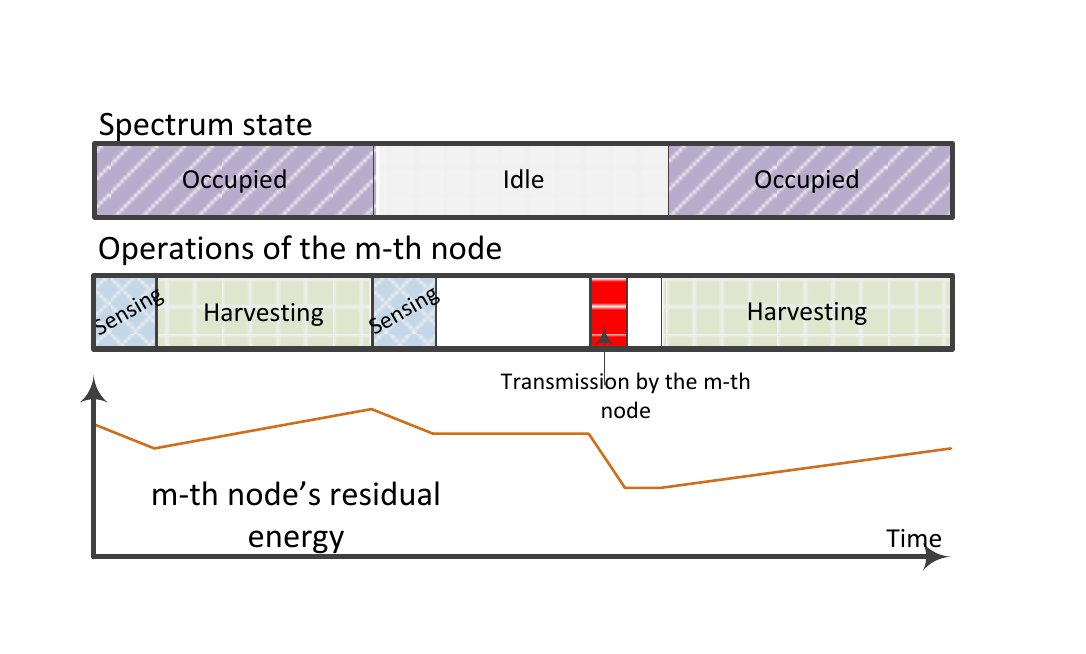}
\par\end{centering}

\caption{\label{fig:park2012optimal}Spectrum access with energy harvesting as discussed in~\cite{park2012optimal}.}
\end{figure}

In~\cite{park2012optimal}, Park \textit{et al.} designed a dynamic, POMDP-based spectrum access control scheme for energy harvesting WSNs as shown in Figure~\ref{fig:park2012optimal}. The nodes are assumed to be unable to access the spectrum during the harvesting stage. Then, the decisions are based on partial information about the spectrum state (occupied or idle) and available energy. The reward of spectrum access, i.e., data transmission, is measured from an acknowledgment message from the data sink which is assumed to be error-free. Similarly, Kashef and Ephremides~\cite{kashef2012optimal} discussed WSN's operation under time varying channels and energy harvesting. The channel access is determined using an MDP policy based on the channel information and the current energy level. The channel state information is known from the feedback from the destination node. The reward function is a discounted sum of the packet delivery. Moreover, an upper bound of the number of successful transmitted packets is derived. Even though the authors of~\cite{park2012optimal,kashef2012optimal} did not directly consider energy harvesting in the reward functions, energy harvesting still affects the future reward values as collecting more energy increases the successful packet delivery.

In a similar context to~\cite{park2012optimal}, Iannello \textit{et al.}~\cite{iannello2012optimality} considered the spectrum access scheduling problem in energy harvesting WSNs. It is assumed that the number of nodes, which are equipped with energy harvesting capability and able to transmit data to a single collector, is larger than the number of available channels. Moreover, to minimize the local data exchange among the centralized scheduling controller and transmitting nodes, the scheduling controller is assumed to have no information about the battery levels of the transmitting nodes. The problem is modeled as a POMDP, and the resulting policy is for the spectrum access of the nodes. The scheduling controller builds its model beliefs by observing the past transmission results as well as the charging and discharging behavior of the batteries. 

Several future research ideas can be inspired from \cite{ park2012optimal, kashef2012optimal,iannello2012optimality} to achieve dynamic spectrum management in energy harvesting WSNs. For example, an upper bound constraint of nodes' waiting time can be imposed and solved using CMDPs. This enables a fair and delay-bounded spectrum access for all nodes. Another potential idea is using stochastic games for non-cooperative algorithms which could reduce the data exchange among nodes.

Different from the above work, Mohamed \textit{et al.}~\cite{mohamed2013adaptive} presented an adaptive data gathering scheme for WSNs with energy harvesting devices. The scheme considers a balance between lossy data compression and the energy budget of the sensors. Most lossy data compression methods can adjust a compression ratio. For example, a higher compression ratio results in poorer data reconstruction performance but more energy savings as less data is transmitted. The MDP model is formulated by incorporating current available energy, expected energy harvesting in the current interval, event occurrence in previous interval, and queued data in the node's buffer. The intended compression error (i.e., error radius between source signal and recovered one at the controller) can be chosen as the MDP's actions. The lower error configuration requires less compression and more energy consumption for data aggregation. Using real-world samples of water pressure and solar energy harvesting data, the simulation shows that the adaptive compression policy provides a small signal reconstruction error at any time during the day or night. With a similar idea, Michelusi \textit{et al.}~\cite{michelusi2013transmission} modeled the ambient energy harvesting in a WSN that collects data of different importance with respect to the system operations. The data importance depends on each sample indication of an event existence, e.g., a high temperature reading indicates a fire event. The ambient energy harvesting is modeled as two states, i.e., good and bad modes. The proposed model enables a node with the bad energy harvesting state to balance between the transmission cost and the energy budget. As a result, a stable overall system service is achieved. In short, in each iteration, the system uses the energy harvesting efficiency, data importance, and current energy budget to predict and take an action so that battery overflow and drainage can be avoided. The MDP's optimal policy is obtained using the policy iteration algorithm.

In~\cite{aprem2013transmit}, Aprem \textit{et al.} considered error control based on ARQ for data retransmission in energy harvesting WSNs. In ARQ, the destination acknowledges a successful packet reception to the sender. Otherwise, the sender assumes an unsuccessful transmission after a timeout period. In the propoesd scheme, a packet acknowledgment (either positive or negative feedback), which is sent back to the transmitter, can be used to build the transmitter's belief and observations about the channel condition and its state information. The states are node's available energy, channel state, number of transmitted packet within a frame, and packet acknowledgment state. The generated beliefs are utilized in the POMDP model to find a near-optimal and low-complexity retransmission policy. 

Niyato and Wang~\cite{niyato2013delay} addressed the stochastic wireless energy harvesting of a mobile node. Under the hard delay requirement, the collected data is received, stored, and forwarded by the mobile node to the destination within a specified threshold constraint. Otherwise, the data, which misses the threshold, will be discarded and removed from the buffer. Therefore, the proposed scheme ensures the delay quality of service requirement given the uncertainty in energy harvesting that are introduced by node mobility. The problem is formulated as a CMDP with delay stages, energy budget levels, and location as the states. The optimal CMDP policy decides whether data transmission is advantageous over the waiting action. 

In many locations, solar energy is considered the most practical source for WSN recharging~\cite{khan2014energy}. Murtaza and Tahir~\cite{murtaza2013optimal} used an MDP to model the battery charging of nodes from solar panels. Accordingly, the energy harvesting is determined by the weather condition, e.g., sunny or cloudy and time of day. The proposed scheme considers the energy requirement of the node at different data transmission rates. Thus, the scheme optimizes the tradeoff between energy harvesting process and the energy consumption. The data collection and data transmission are assumed to follow the Poisson distribution. The node decides whether data transmission is required for event detection using the policy obtained from the MDP. Similarly, Mao \textit{et al.}~\cite{mao2014joint} considered the problem of maximizing the amount of transmitted data in energy harvesting WSNs. Data transmission may be deferred because of various reasons including a drained battery, an empty transmission buffer, and poor transmitting channel condition. The energy harvesting and allocation problem is formulated as an MDP which is later solved using the value iteration method. The data receiver notifies the transmitter about the CSI. The infinite time horizon MDP has the state as the node's available energy, data buffer, harvesting state, and channel state. The actions specify energy allocation for transmission and sensing operations. 

Then, Nourian \textit{et al.}~\cite{nourian2014optimal} designed a transmission scheme over an error-prone channel in energy harvesting WSNs. The channel's data dropping depends on the transmit power, and the channel gain and fading properties. This dropping problem affects the data acknowledgment from the receiver to the transmitting node, i.e., an imperfect and incomplete feedback. An MDP is used to minimize the average error from the channel. To calculate the channel's average error, a Kalman filter-based channel estimation technique is used, see~\cite{barkat2005signal} for an introduction to the Kalman filter. The MDP is solved using dynamic programming, and the suboptimal solution is obtained with reduced computational complexity. In a similar application, Ren \textit{et al.}~\cite{ren2014dynamic} addressed the scheduling and activation problem of rechargeable nodes in event monitoring WSNs. Monitored events and node recharging processes are assumed to be random. Firstly, it is assumed that a node has full information about the event occurrence from the previous iteration. The problem is formulated as an MDP. Herein, energy budget, node's activation history, and event occurrence history are the states of the MDP. Secondly, when the node has partial information about the events (knowledge about currently active events), the problem is formulated and solved as a POMDP. In this case, the energy budget, node activation history, and node beliefs about event occurrence are all used for POMDP's states initialization. Furthermore, cooperative event detection by multiple nodes is also discussed. In a Matlab-based simulation, the event occurrence is assumed to follow a Weibull or a Pareto distribution. The results show that the activation policy captures events with higher probabilities as the battery capacity increases.

\subsection{Dynamic Optimization and Resource Allocation}

WSNs operate in dynamic environments, and sensor nodes need to adapt to the changes to minimize their resource consumption. For example, a node that optimizes its channel access protocol to a congestion condition can minimize its overall energy consumption. Table~\ref{tab:dynamic_optimization} outlines dynamic optimization methods that are based on MDP schemes.

\begin{table*}
\caption{\label{tab:dynamic_optimization}Summary of the surveyed dynamic optimization methods (DSE = design space exploration, E2E = end-to-end).}

\begin{centering}
\begin{tabular}{|c|>{\centering}m{1.1cm}|>{\centering}m{1cm}|>{\centering}m{1.2cm}|>{\centering}m{3cm}|>{\centering}m{3cm}|>{\centering}m{2.6cm}|}
\hline 
\textbf{\noun{Application context}} & \textbf{\noun{Article}} & \textbf{\noun{Type}} & \textbf{\noun{Decision}} & \textbf{\noun{States}} & \textbf{\noun{Actions}} & \textbf{\noun{Rewards/costs}}\tabularnewline
\hline 
\hline 
Task scheduler &~\cite{zhu2007tasks} & MDP & Centralized & Executed tasks & Allocate time slots to tasks & Finished tasks, missed deadlines, energy\tabularnewline
\hline 
System maintenance &~\cite{misra2010probabilistic} & MDP & Centralized & Exhausted nodes & Replace (or keep) an exhausted node  & Deployment cost, network performance\tabularnewline
\hline 
\multirow{4}{*}{Dynamic configuration} &~\cite{grassi2012knowledge} & MDP & Centralized & DSE's hardware components & Modify hardware components & Network performance, components cost\tabularnewline
\cline{2-7} 
 &~\cite{munir2012mdp} & MDP & Centralized & Hardware components & Modify hardware components & Network performance, components cost\tabularnewline
\cline{2-7} 
 &~\cite{kovacs2012mixed} & MDP, POMDP & Distributed & Position of generated data & Sleep, inferior , sample, or aggregate. & E2E delay, energy, data consistency\tabularnewline
\cline{2-7} 
 &~\cite{lin2013markovian} & MDP & Centralized & Active links, buffer occupancies, and available energies & Join the transmission set  & Energy\tabularnewline
\hline 
\end{tabular}
\par\end{centering}

\end{table*}

\subsubsection{Task scheduler}

Zhu \textit{et al.}~\cite{zhu2007tasks} discussed task scheduling and allocation of parallel applications in heterogeneous WSNs. This scheduling process considers the energy consumption of the heterogeneous nodes and parallel tasks' deadlines. For example, a resourceful node can finish the task in shorter time but it consumes more energy than that of a less resourceful node. Considering the task dependencies, the scheduler uses an MDP framework to make the scheduling decision and assign each task to the suitable nodes. The states include the currently executed tasks and task allocation over heterogeneous nodes. The action space corresponds to time slot allocation of tasks to the available nodes. Moreover, the reward function evaluates the task release (finishing) time, missed deadlines, and energy consumption during task execution. The MDP-based task allocation method is compared with a heuristic method and a greedy one. The heuristic policy considers the task's release time, while the greedy policy considers the energy consumption. The greedy policy does not guarantee tasks' deadlines, and the MDP-based task allocation leads to less energy consumption than that of the heuristic policy.

\subsubsection{System Maintenance}

Misra \textit{et al.}~\cite{misra2010probabilistic} suggested an algorithm for modeling WSN maintenance. In particular, the designed algorithm considers the tradeoff between node replacement and network performance. The MDP policy decides the minimum number of nodes that must be replaced to maintain the network operation, and therefore minimizes the network's operational cost. Equally important, the algorithm takes into account the replacement cost per sensor, e.g., when replacing more sensors, the cost per sensor decreases. The states are defined as the number of drained nodes in the network. Additionally, a maintenance action is defined as replacing a specific set of nodes at each maintenance instance.

\subsubsection{Dynamic Configuration and Lifetime Modeling}

In~\cite{grassi2012knowledge}, Grassi \textit{et al.} considered the computer-aided design (CAD) of WSNs that helps in selecting the optimized configuration of hardware components. The node's components, such as the central processing unit (CPU), memories, and radio transceiver, are designed to fit the deployment scenario and requirements. An MDP is used instead of the conventional methods which require complex simulation analysis of design space exploration (DSE). The MDP's states characterize different design solutions of the DSE problem. The actions describe the component changes that can be applied to each solution and result in transition to a new solution state. In the same context, Munir \textit{et al.}~\cite{munir2012mdp} suggested tuning the node's configuration using an MDP. For example, the sampling frequency of the node is optimized to match the responsiveness requirement and environment condition. The full list of system's parameters, such as CPU's voltage, frequency, sampling rate, defines the MDP's states.

Another direction for system configuration is to optimize nodes' run time operations to match the dynamic environment conditions. For example, Kovacs \textit{et al.}~\cite{kovacs2012mixed} introduced a methodology for dynamically optimizing WSN protocols such as routing, data aggregation, and topology control. Essentially, the considered performance metrics include data gathering delay, energy consumption, and data consistency. The actions are switch to idle mode, listen to events, sample readings, and aggregate packets. Likewise, Lin \textit{et al.}~\cite{lin2013markovian} addressed the multi-hop transmission in both cooperative and non-cooperative WSNs at MAC, routing, and physical layers. In cooperative networks (CTNs), sensor nodes can decide to cooperate for creating a virtual multiple-input multiple-output (VMISO) link that is useful for delivering data to a sink at a distance. These cooperating nodes (i.e., a co-operator) are called the transmission set and each of them is assumed to have data in its transmission queue. The analysis assumes that no neighbor nodes can transmit at the same time, and hence hidden terminals can cause collisions. The states include the transmission nodes, buffer sizes, and available energies. Experimental results reveal that the CTN with one co-operator extends the non-CTN's lifetime by a factor of 1.89. The network's lifetime is also linearly proportional to the battery capacity by factors of 1, 1.6, and 2.1 in non-CTNs, CTNs with 2 co-operators, and CTNs with 1 co-operator, respectively.

In summary, these algorithms for dynamic configuration and lifetime modeling could be particularly challenging in outdoor and harsh environments, where changing weather conditions influence the wireless channel and interference models.

\subsection{Duty Cycling and Medium Access Control (MAC)}

WSNs operate under limited energy resource and the simultaneous activation of all autonomous nodes can ineffectually waste this limited energy budget~\cite{wang2006survey}. For example, continuous activation of all sensors attached to the human body in activity recognition applications is not energy friendly. Moreover, centralized activation systems require energy expensive data exchange among network components. Duty cycling is the mechanism to manage the active and sleep modes of nodes while performing the required operations. MDPs are used to optimize duty cycle and MAC as shown in Table~\ref{tab: duty_cycle}.

\begin{table*}
\caption{\label{tab: duty_cycle}Summary of duty cycle and MAC protocols (SG = stochastic game, SNR = signal to noise ratio, E2E = end-to-end, CAP = contention access period, CFP = contention free period).}

\begin{centering}
\begin{tabular}{|c|>{\centering}m{1.1cm}|>{\centering}m{1cm}|>{\centering}m{1.2cm}|>{\centering}m{3cm}|>{\centering}m{3cm}|>{\centering}m{2.7cm}|}
\hline 
\textbf{\noun{Application context}} & \textbf{\noun{Article}} & \textbf{\noun{Type}} & \textbf{\noun{Decision}} & \textbf{\noun{States}} & \textbf{\noun{Actions}} & \textbf{\noun{Rewards/costs}}\tabularnewline
\hline 
\hline 
\multirow{1}{*}{Duty cycle management} &~\cite{yuan2011balanced} & MDP & Distributed & Initialized, sleep, active, dead & Change node mode & Energy\tabularnewline
\hline 
\multirow{6}{*}{MAC} &~\cite{zhao2008using} & SG & Distributed & Number of opponent nodes & Transmit, listen, or sleep & Collision, energy\tabularnewline
\cline{2-7} 
 &~\cite{wang2009energy} & MDP & Distributed & SNR & Select access time & Collision\tabularnewline
\cline{2-7} 
 &~\cite{jagannath2013hybrid} & MDP & Centralized & Buffer occupancy, available energy & Select transmission slots & Energy, buffer cost, failing-penalty\tabularnewline
\cline{2-7} 
 &~\cite{mehta2010energy} & SG & Distributed & Number of competing nodes & Transmit, listen, or sleep & Energy, delay\tabularnewline
\cline{2-7} 
 &~\cite{shrestha2014distributed} & MDP & Centralized & Buffer occupancies in super-frames & Transmit (CAP, CFP, or both), or wait & Energy, throughput, bandwidth\tabularnewline
\cline{2-7} 
 &~\cite{pajarinen2014optimizing} & DEC-POMDP & Distributed & Buffer occupancies, traffic source & Transmit or listen & Throughout, E2E delay\tabularnewline
\hline 
Spectrum access &~\cite{seokwon2012energy} & POMDP & Centralized & Spectrum occupancy & Sense (or occupy) the spectrum & Ultra low power networks\tabularnewline
\hline 
\end{tabular}
\par\end{centering}

\end{table*}

\subsubsection{Duty Cycle Management}

Yuan \textit{et al.}~\cite{yuan2011balanced} proposed the duty cycling algorithm for WSNs based on an MDP. The available energy is the main parameter to decide on the activation of the sensor nodes. In particular, the algorithm guarantees that the set of active nodes consists of the connected nodes with the highest energy budgets. The MDP's states correspond to node's states of initialized, sleep, active, or dead modes. Each node must broadcast its available energy to other nodes, and therefore full information is available for nodes. The key result is that the energy conservation is inversely proportional to the number of connected neighbors.

\subsubsection{Media Access Control (MAC)}

Zhao \textit{et al.}~\cite{zhao2008using} suggested a MAC protocol by using a stochastic game, where each node deals with other nodes as opponents. The MAC operation is divided into cycles and each cycle interval is for a packet transmission. In each interval, a node takes an action of: i) transmitting a buffered packet, (ii) switching to listen mode, or (iii) switching to sleep mode. Moreover, the nodes dynamically optimize their MAC contention parameters, e.g., backoff time, based on the channel condition. This distributed algorithm does not require exchanging action information among nodes. Instead, the other nodes' actions are predicted using the historical observation. In particular, the detection of competing nodes considers various cross-layer parameters such as SNR, transmission probability, collision probability, and datagram loss ratio (DLR). Accordingly, the current state is predicted as the number of opponent nodes in each interval.

\begin{figure}
\begin{centering}
\includegraphics[width=1\columnwidth,trim=1cm 1cm 1cm 0.5cm]{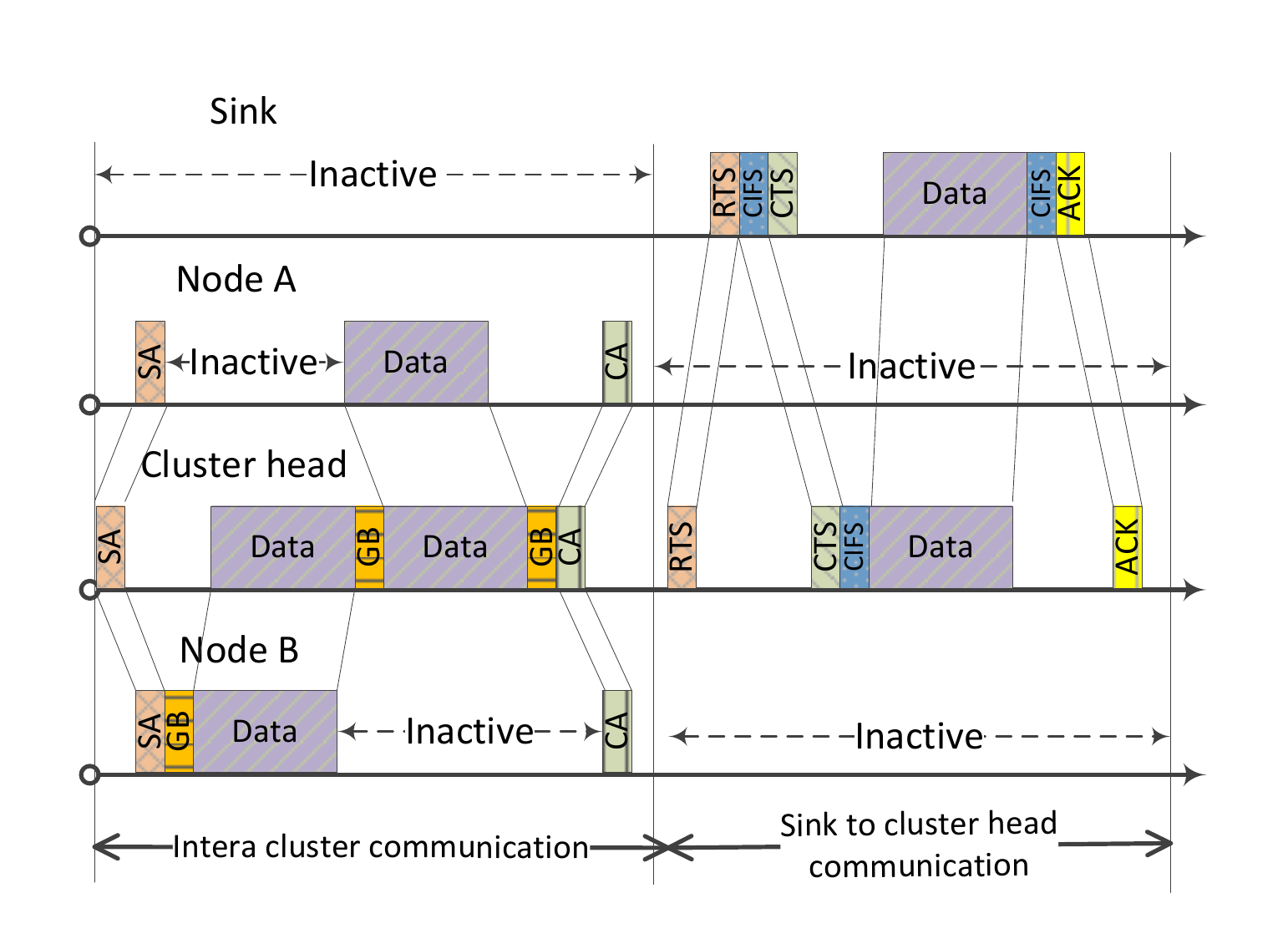}
\par\end{centering}

\caption{\label{fig:jagannath2013hybrid}Timing and data exchange among a cluster head, two sensor nodes and a sink using the hybrid MAC protocol proposed
in~\cite{jagannath2013hybrid} (CIFS = contention window inter-frame spacing, RTS = request to send, CTS = clear to send, ACK = acknowledge, SA = slot announcement, GB = guard-band, CA = cumulative acknowledgment).}
\end{figure}

In~\cite{wang2009energy}, Wang \textit{et al.} suggested an enhancement to the carrier sense multiple access with collision avoidance (CSMA/CA) protocol in WSNs. Basically, the study analyzes CSMA/CA and its limitations in slowly fading Rayleigh channels. The Rayleigh channel is modeled as an MDP to predict the channel fading state. The SNR is quantized into ranges to represent channel fading, and the node decides its channel access time based on the channel state. Assuming that the channel state can only change to one of the two neighbor states, the transmission matrix is a tridiagonal matrix, i.e., a matrix with zero entries except for main diagonal, and one line above and below the main diagonal. This tridiagonal form helps in determining the state at future time slots without specifying the initial state. In a similar context, Jagannath \textit{et al.}~\cite{jagannath2013hybrid} introduced a MAC protocol that considers the physical layer parameters for optimizing scheduling decisions. The protocol is designed for underwater WSNs where nodes' battery replacement is a laborious task. Two MAC protocols are used: TDMA protocol for intra-cluster transmission and CSMA/CA for inter-cluster transmission, i.e., cluster heads and sink data exchange. The exchanged data and control messages are shown in Figure~\ref{fig:jagannath2013hybrid}. CSMA/CA's control messages include contention window inter-frame spacing (CIFS), request to send (RTS) and clear to send (CTS) handshaking, and acknowledge message (ACK). By contrast, TDMA uses coordinating messages such as slot announcement (SA), guard-band (GB), and cumulative acknowledgment (CA) packets. Within each cluster, a node selfishly estimates its required transmission allocation using an MDP model and sends the estimation to the cluster head. Then, the cluster head, based on the channel quality and the data priority, assigns the MAC's slots to transmitting nodes to minimize the energy consumption. The state of the node is a buffer size and battery state. Then, the MDP's action is the number of slots that the node requires for transmission. The reward is composed of the energy consumption in data transmission, buffering cost (avoid buffer overflow and hence data loss), node failure, and energy saving in sleep mode.

Similar to~\cite{zhao2008using}, Mehta \textit{et al.}~\cite{mehta2010energy} proposed a suboptimal backoff algorithm for a MAC protocol to avoid collision in WSNs. The backoff algorithm is used in CSMA/CA and is decided by each node based on the transmission behavior of the other nodes. The players in the stochastic game are the sensor nodes competing for channel access. The actions are transmit, listen, or sleep. Furthermore, each node tunes its contention window size during the transmit mode. The proposed algorithm considers the energy consumption, transmission delay, and throughput. The proposed backoff algorithm is validated using a Matlab-based simulation with 100 nodes. The conventional MAC algorithm achieves high packet transmission rates for small numbers of nodes. However, the rate decreases as the number of nodes increases, e.g., more than 20 nodes. The proposed backoff algorithm enhances the scalability of conventional MAC protocols by achieving better performance at the increased number of nodes.

Unlike ~\cite{jagannath2013hybrid} which considers a hybrid CSMA/CA and TDMA protocols at different hierarchical levels, Shrestha \textit{et al.}~\cite{shrestha2014distributed} divide the channel access into two periods of contention (CSMA/CA) and contention-free (TDMA) phases. The proposed design is for tackling the problem of poor CSMA/CA's performance (i.e., energy consumption and throughput) when the channel is congested. This hybrid protocol is adopted in IEEE 802.15.4 networks when the nodes encounter large buffer sizes. A large buffer size is an indicator of a congested channel, and data is dropped if the maximum buffer size is exceeded. Based on the buffer occupancy, the infinite time horizon MDP model is formulated and solved to obtain the transmission policy: transmit in contention access period (CAP), transmit in contention free period (CFP), transmit in both CAP and CFP, or continue waiting without transmission. The reward is composed of energy consumption, required bandwidth, and throughput. 

Apart from the aforementioned work, Pajarinen \textit{et al.}~\cite{pajarinen2014optimizing} cast the problem of medium access using as a DEC-POMDP to capture tempo-spatial correlation in data traffic. This MAC protocol is designed to consider the tradeoff between high throughput and small delay. The DEC-POMDP model is employed because of sensors' noise and partial information about other transmission. Each transmitting node builds its belief about others' transmissions by monitoring the interference level, and therefore the protocol does not require control data exchange among nodes. The system states include two parameters: traffic source data generation (data and no-data generated), and the current buffer occupancy of the transmission controllers.

On the negative side, applying an offline solution to find an optimal MAC policy requires disseminating a new policy when there are changes in the network, which would be relatively costly. Moreover, even though the stochastic games are well suited for MAC management, the computational complexity becomes critical in large scale WSNs.

\subsubsection{Spectrum Access}

In~\cite{seokwon2012energy}, Seokwon \textit{et al.} considered the spectrum access of multiple WSNs with ultra low power devices operating simultaneously. This introduces interference and significant energy consumption. Hence, the study proposes using a POMDP for spectrum access decisions which reduces switching among transmitting channels. The POMDP's states include the spectrum occupancy state, and the action space consists of commands to sense or occupy the spectrum. However, it is found that the transmission overhead can be considerable when sending the spectrum access schedule from the central coordinator to each sensor at the beginning of each transmission cycle.

As demonstrated with several examples in this section, implementing resource and power optimization algorithms using MDPs is possible and can significantly improve WSN operations. Sensing coverage and object detection are other important issues in the development of WSNs. In the following section, we review the existing literature on MDP-based sensing coverage and object detection algorithms in WSNs.

\section{Sensing Coverage and Object Detection}\label{sec:Sensor-coverage}

Sensor nodes can be deployed manually or randomly. Moreover, some nodes can be mobile, and thus the deployed locations can change dynamically. In all deployment scenarios, MDPs are used to achieve the following benefits: 
\begin{itemize} 
\item \emph{Sensing coverage}: The MDP models are used to predict the minimum number of active nodes to achieve the required coverage performance over time. Moreover, some work assumes that mobile nodes can change their location. In the latter, the MDP can predict optimal movement strategies (e.g., movement steps and directions) of the nodes to cover the monitored area. 
\item \emph{Target tracking}: To increase the probability of object detection, WSNs use MDPs to predict the future locations of the targeted object, and to activate sensors at the expected locations and switch off sensors in other locations. Additionally, the MDP models can predict optimal movement directions of nodes to increase the probability of target detection.
\end{itemize}

\subsection{Sensing Coverage and Phenomenon Monitoring}

Sensing coverage describes the ability of sensor networks to provide complete information about the monitored area. The sensor coverage problem is coupled with other networking and connectivity perspectives of WSNs~\cite{ghosh2008coverage,wang2011coverage}. For example, although some nodes may not perform reading, they have to be active to relay sensed data to a sink. Table~\ref{tab:sensor_coverage} outlines notable studies of sensor coverage modeling using MDPs. For a clear discussion of these methods, we define three terms that are widely used in the literature. 
\begin{itemize} 
\item \emph{Area of interest (AoI)}: AoI is the area that must be precisely covered over time. Subareas inside the AoI can be spatially correlated with each other, and therefore using correct models enables predicting phenomena at uncovered part based on other covered subareas. Moreover, one specific area's readings can be temporally correlated, which means that the future readings can be predicted from the past ones. 
\item \emph{Points of interest (PoI)}: PoI reflects the interest of phenomena readings at specific location. Again, location points can be temporally and spatially correlated, and hence can be extracted from each other. 
\item \emph{Detection probability}: In object tracking, the detection probability describes the level of certainty about an object's location that can be achieved by activating a set of nodes. Accordingly, when a higher detection probability is required, generally more active sensors are needed.
\end{itemize}

\begin{table*}
\caption{\label{tab:sensor_coverage}Summary of surveyed sensing coverage applications (PTZ = pan\textendash{}tilt\textendash{}zoom, AOI = area of interest, E2E = end-to-end).}

\centering{}%
\begin{tabular}{|c|>{\centering}m{1.1cm}|>{\centering}m{1cm}|>{\centering}m{1.2cm}|>{\centering}m{3cm}|>{\centering}m{3cm}|>{\centering}m{2.6cm}|}
\hline 
\textbf{\noun{Application context}} & \textbf{\noun{Article}} & \textbf{\noun{Type}} & \textbf{\noun{Decision}} & \textbf{\noun{States}} & \textbf{\noun{Actions}} & \textbf{\noun{Rewards/costs}}\tabularnewline
\hline 
\hline 
\multirow{3}{*}{Object detection} &~\cite{fei2010pomdp} & POMDP & Centralized & Sensor activations & Select $k$ active nodes & Detection probability and coverage\tabularnewline
\cline{2-7} 
 &~\cite{ota2012oracle} & MDP & Distributed & Increase or decrease the detection probability & Move the actor node & Detection probability\tabularnewline
\cline{2-7} 
 &~\cite{vaisenberg2014scheduling} & POMDP & Distributed & Zoomed-in or zoomed-out camera & Manage PTZ camera zooming and direction & Detection probability\tabularnewline
\hline 
\multirow{2}{*}{Rescue applications} &~\cite{murtaza2013priority} & POMDP & Distributed & Cells within an AOI & Select a moving direction & cell coverage\tabularnewline
\cline{2-7} 
 &~\cite{mondal2012optimal} & MDP & Centralized & Distance between the coordinator and previous deployed relay  & Choose movement steps & Connectivity, E2E delay, deployment cost\tabularnewline
\hline 
\end{tabular}
\end{table*}

\subsubsection{Object Detection}

The connected $k$-coverage problem is a common formulation of the coverage problem, where $k$ connected nodes must be active at any time instant. Therefore, the problem formulation insures coverage quality of the network. Fei \textit{et al.}~\cite{fei2010pomdp} addressed the problem of enhancing the area coverage in WSNs. Assuming a dense sensor deployment, the algorithm selects the most useful sensors to be active. Therefore, the other sensors can switch to sleep mode to conserve their energy. Assuming a network that consists of $n$ nodes, an action is taken to activate $k$ out of the $n$ sensor nodes at each decision interval. However, without complete information about the targeted object, the algorithm is designed based on an POMDP, and hence the object's location is probabilistically identified. The reward function is increased by one unit if the object moves within an active sensor detection range, and a negative reward is received otherwise. In~\cite{ota2012oracle}, Ota \textit{et al.} presented an optimized, MDP-based event detection mechanism by using mobile sensor nodes. The mechanism is to minimize mobile robot's (also called actor node) movement strategy, while maximizing the event detection probability. The parameters of the model are predicted using maximum-likelihood estimation (MLE), see~\cite{barkat2005signal} for an introduction to the MLE. The states are structured to capture an improvement or deterioration of the detection probability, i.e., the state is either ``increase'' or ``decrease'' in detection probability. The MDP model is solved using reinforcement learning algorithms.

\begin{figure}
\begin{centering}
\includegraphics[width=0.95\columnwidth,trim=1cm 1cm 1cm 0.5cm]{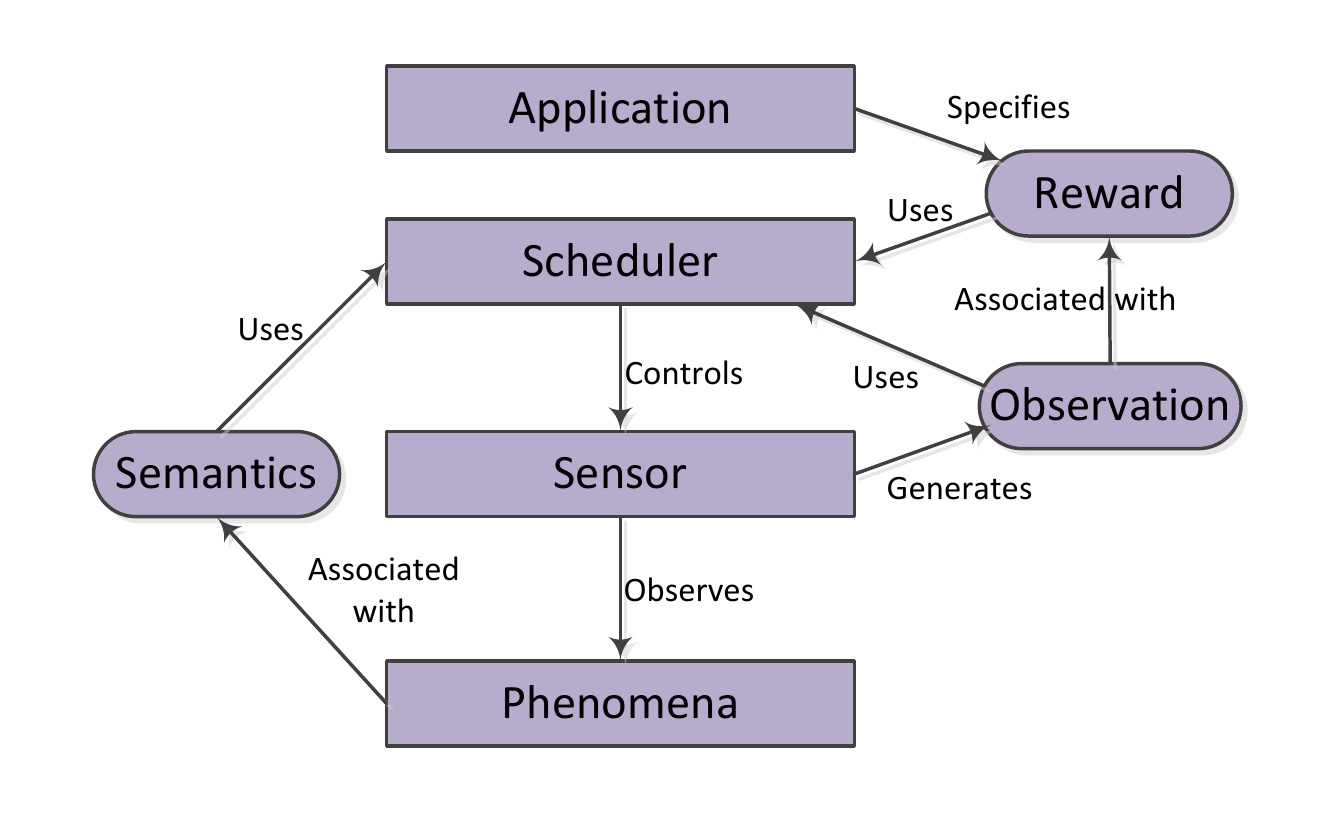}
\par\end{centering}

\caption{\label{fig:vaisenberg2014scheduling} The block diagram of the scheduling framework presented in~\cite{vaisenberg2014scheduling}.}
\end{figure}

Vaisenberg \textit{et al.}~\cite{vaisenberg2014scheduling} utilized a POMDP to model the future physical phenomenon. Consequently, the AoI can be better monitored and covered by the deployed monitoring system. The remotely-sensed values are considered as the POMDP's states. For example, consider a pan\textendash{}tilt\textendash{}zoom (PTZ) camera monitoring system as a potential application. The designed system optimizes camera directions and zooming actions to maximize event detection probabilities. A ``zoomed-in'' action help capture images with high resolution, but with small AoI. On contrary, a ``zoomed-out'' action provides images for a larger AoI. Then, the rewards are increased when objects are within the captured images, e.g., object occurrence can be recognized by an image processing technique. Figure~\ref{fig:vaisenberg2014scheduling} shows the block diagram of the developed system. The proposed decision policy is evaluated for a human monitoring system and compared with other standard methods such as a round robin-based method that continuously cycles the camera's focus between zooming in and out. The proposed system outperforms the other standard methods and gain the highest total reward values.

\subsubsection{Area Coverage in Rescue Applications}

Murtaza \textit{et al.}~\cite{murtaza2013priority} discussed the coverage perspectives of using WSNs for path planning of victim evacuation from disaster areas. The path planning aims to determine the optimal paths for short-time rescuing operation, which is critical for saving human lives. Assuming unknown number and locations of victims, a POMDP model locates casualties with the shortest possible time. Moreover, due to the disaster damages, the mobile robot has incomplete information about the covered area's terrains and how the casualties are distributed throughout the area. Therefore, the proposed solution cannot prioritize subareas of the total AoI. The states correspond to searching squares of the AoI's grid. Correspondingly, the actions are the eight possible moving directions to neighboring squares. A robot will acknowledge the base station if it can find a victim in any locations during its movement. Therefore, the rescue team updates its belief map simultaneously. Moreover, the probability of finding other victims in nearby locations is also increased. Otherwise, if no case is found in the scanned square, a clear message is also reported.

In a similar context, Mondal \textit{et al.}~\cite{mondal2012optimal} discussed the optimal deployment of relay sensors in emergency scenarios and without prior knowledge of terrains. It is to decide sensor placement for maintaining good connectivity, e.g., small end-to-end delay at low cost. The problem is modeled as an MDP. A coordinator, which deploys relay nodes, moves through the AoI and decides whether a relay is needed at each step. The distance between the coordinator and the last deployed relay is considered as the current system state. The numerical results consider a corridor area scenario with a restricted number of available relays and show that deploying more relays decreases the total energy consumptions of the network.

In conclusion, using MDPs for area coverage in rescue applications, as implemented in~\cite{murtaza2013priority,mondal2012optimal}, is an interesting and useful idea to save human lives. However, more experimental validation within practical environments should be conducted before using these systems in real rescue cases.

\subsection{Target Tracking and Localization}

The object tracking component is an important part of WSNs in monitoring and surveillance applications. The core object classification and detection process can be efficiently performed by supervised machine learning algorithms~\cite{alsheikh2014machine}. Conversely, this section explains energy efficiency aspects of tracking applications which can be modeled as MDPs, e.g., minimum node activation. The MDP-based methods analyze the tradeoff between the energy consumption and the object detection accuracy. Additionally, they predict the next object activity and location that can be used to trigger the required actions such as sensor and alarm activation. A comparison of these target tracking methods is presented in Table~\ref{tab:target_tracking}. In column ``Parameters'', the detection accuracy is usually given by the probability of false alarm generated by the algorithm.

\begin{table*}
\caption{\label{tab:target_tracking}Summary of target tracking and localization methods in WSNs (SG = stochastic game, HMDP = hierarchical Markov decision process, CH = cluster head, CMDP = constrained Markov decision process).}

\centering{}%
\begin{tabular}{|c|>{\centering}m{1.1cm}|>{\centering}m{1cm}|>{\centering}m{1.2cm}|>{\centering}m{3cm}|>{\centering}m{3cm}|>{\centering}m{2.6cm}|}
\hline 
\textbf{\noun{Application context}} & \textbf{\noun{Article}} & \textbf{\noun{Type}} & \textbf{\noun{Decision}} & \textbf{\noun{States}} & \textbf{\noun{Actions}} & \textbf{\noun{Rewards/costs}}\tabularnewline
\hline 
\hline 
\multirow{5}{*}{Cooperative object tracking} &~\cite{fuemmeler2008smart} & POMDP & Centralized & Node activations (sleep, active) & \multirow{7}{3cm}{\centering{}Select active nodes} & Energy, detection probability\tabularnewline
\cline{2-5} \cline{7-7} 
 &~\cite{zhan2010active} & MDP & Centralized & Estimated Adversary's region &  & Energy consumption, detection probability\tabularnewline
\cline{2-5} \cline{7-7} 
 &~\cite{atia2011sensor} & POMDP & Centralized & Estimated object's location &  & Energy consumption, detection probability\tabularnewline
\cline{2-5} \cline{7-7} 
 &~\cite{fuemmeler2011sleep} & POMDP & Centralized & Estimated object's location, sleep times &  & Energy consumption, detection probability\tabularnewline
\cline{2-5} \cline{7-7} 
 &~\cite{huang2010distributed} & SG & Centralized & Quantized spectrum bandwidth &  & Energy, successful transmission\tabularnewline
\cline{1-5} \cline{7-7} 
\multirow{4}{*}{Clustered tracking systems} &~\cite{yeow2007energy} & HMDP & Distributed (CH) & CH's state (sensing, listening, or tracking) &  & Sensing rate, detection probability\tabularnewline
\cline{2-5} \cline{7-7} 
 &~\cite{misra2012localized} & MDP & Distributed (CH) & Sensor's state (sleep, fully or partially active) &  & Energy, detection probability\tabularnewline
\cline{2-7} 
 & \multirow{2}{1.1cm}{\centering{}\cite{jamal2012event}} & \multirow{2}{1cm}{CMDP} & \multirow{2}{1.2cm}{Centralized} & (Lower tier) buffer occupancy, congestion matrix & Select active nodes and detection threshold & Network congestion, detection probability\tabularnewline
\cline{5-7} 
 &  &  &  & (Upper tier) priority matrix, competing users & Assign a spectrum & Priority\tabularnewline
\hline 
\multirow{2}{*}{Multiple target tracking} &~\cite{li2009approximate} & POMDP & Centralized & Targets' locations, node activations  & \multirow{4}{3cm}{\centering{}Select active nodes} & Energy consumption, detection probability\tabularnewline
\cline{2-5} \cline{7-7} 
 &~\cite{zhang2013uts} & POMDP & Centralized & Targets' locations and velocities &  & Nodes' interception risk, detection accuracy\tabularnewline
\cline{1-5} \cline{7-7} 
\multirow{3}{*}{Health and body networks} &~\cite{au2011carer} & POMDP & Centralized & Human body activities &  & Energy consumption, detection probability\tabularnewline
\cline{2-5} \cline{7-7} 
 &~\cite{zois2013unified} & POMDP & Centralized & Human body activities (sit, stand, etc) &  & Energy consumption, detection probability\tabularnewline
\cline{2-7} 
 &~\cite{pietrabissa2013optimal} & MDP & Centralized & Asset's location & Move (north, south, east, or west) & Transportation delay\tabularnewline
\hline 
Prioritized data delivery &~\cite{pino2014selective} & MDP & Distributed & Targets' locations and velocities & Send or discard a message & Detection probability\tabularnewline
\hline 
\end{tabular}
\end{table*}

\subsubsection{Cooperative Object Tracking}

In~\cite{fuemmeler2008smart}, Fuemmeler and Veeravalli proposed a duty cycle management policy for tracking applications in densely deployed WSNs. A few sensor nodes detect an object at the same time. Therefore, the other sensors can be switched to sleep mode without affecting the detection performance. An asleep sensor is assumed to stay in inactive mode until its internal sleep timer finishes, and it cannot be switched on by any external signal from the control unit. There is a minimum threshold for the number of active nodes that must be considered at any time instant. The developed system is based on a POMDP model to optimize the tradeoff between sleep nodes and detection performance using a suboptimal policy. The nodes are assumed to be in one of two states: sleep and active modes. The sensors' sleep decisions are managed by a central unit, which decides the sleep time for each sensor. The cost function is composed of energy saving and a detection performance.

\begin{figure}
\begin{centering}
\includegraphics[width=0.7\columnwidth,trim=1cm 1cm 1cm 0.5cm]{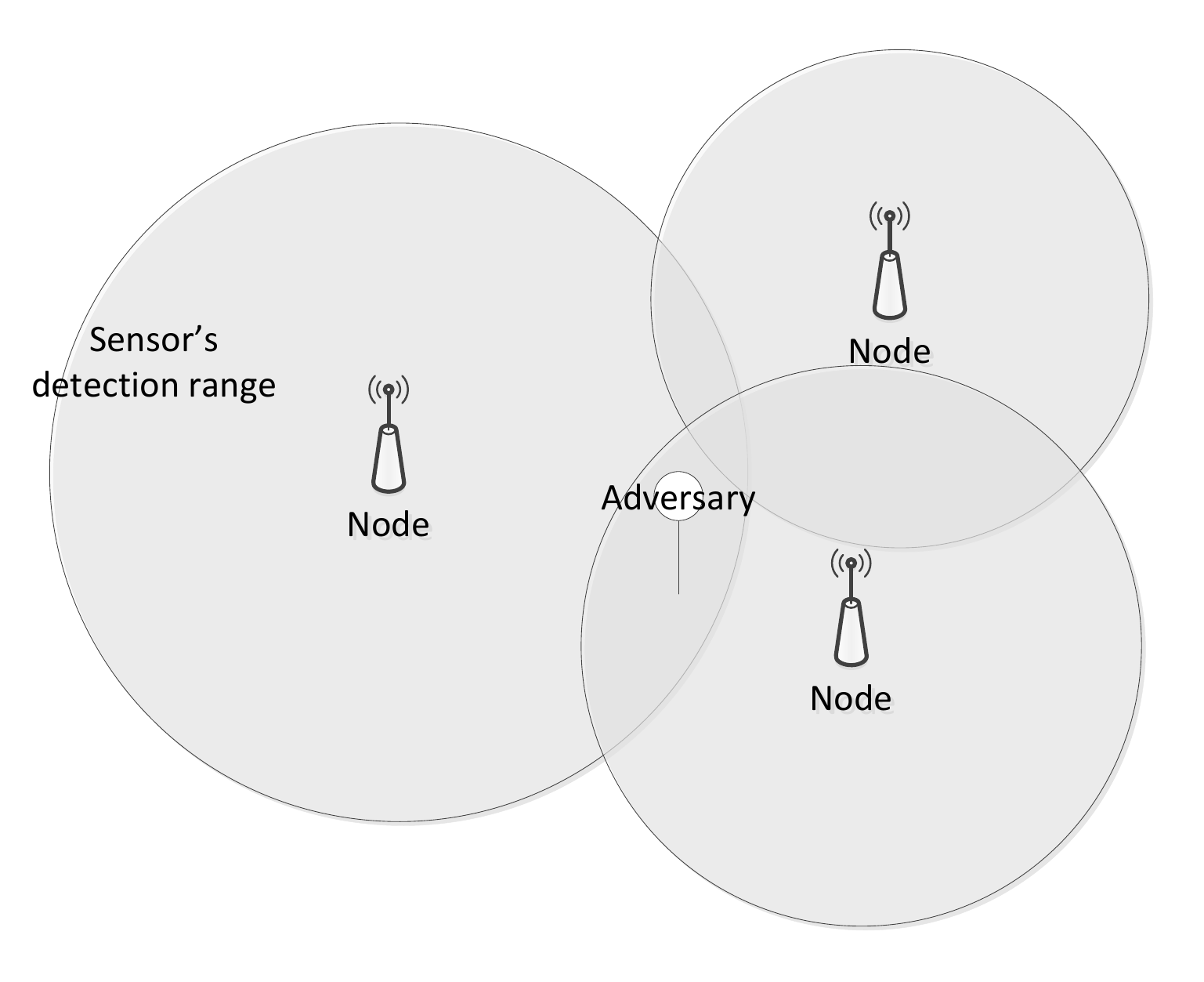}
\par\end{centering}

\caption{\label{fig:zhan2010active} Adversary detection by nodes where each node has a different detection range as presented in~\cite{zhan2010active}.}
\end{figure}

For object detection in security and monitoring application, Zhan and Li~\cite{zhan2010active} proposed the scheme to locate malicious objects in WSNs (Figure~\ref{fig:zhan2010active}). An adversary's location is found by cooperating nodes, and the final location is extracted by an MDP. The MDP's states represent the possible regions surrounding a node, and a region can be at the intersection of nodes' detection areas. Therefore, the policy determines the set of nodes to be activated to maximize the malicious object detection. The simulation of a grid topology indicates that the ratio between the localization error and coverage radius is less than 0.3.

As an extension of the previous studies, Atia \textit{et al.}~\cite{atia2011sensor} considered the problem of object tracking under two sensor deployments: overlapped and non-overlapped sensing ranges. The overlapped case occurs when the targeted object is covered by many sensor ranges, and the non-overlapped one considers object detection by a single active node. In these cases, the energy and detection efficiency tradeoff is optimized using a POMDP. The POMDP's states refer to beliefs about the locations of an object which are stored in a central controller to derive optimal sensor selection process. Later, Fuemmeler \textit{et al.}~\cite{fuemmeler2011sleep} extended the studies by assuming that the sensor locations can be outside the covered areas. Each node can be either in sleep or wake up modes. Therefore, the target object can leave the network area. A centralized controller that uses the POMDP determines the node activation and sleep time. 

Huang \textit{et al.}~\cite{huang2010distributed} considered the problem of object detection, where the channel spectrum is limited and shared among nodes. A node takes actions to control its operation state which is active or sleep. Moreover, a coordinator manages the required spectrum bandwidth by considering the number of active nodes. The joint actions of all nodes are important from two perspectives. Firstly, it is used in spectrum management to decide the transmission. Secondly, it is required to optimize the object detection task by selecting the number of active nodes. The problem is solved using a Q-learning algorithm to find a correlated equilibrium. The experimental analysis considers a 2$\times$2 grid topology and 10 states of the available spectrum bandwidth. The correlated equilibrium policy is found after 300 update iterations, which is relatively fast.

To sum up, the algorithms proposed in~\cite{fuemmeler2008smart,zhan2010active, atia2011sensor,fuemmeler2011sleep,huang2010distributed} require an offline learning phase at a central unit. This centralized design incurs high costs of gathering data to a base station, and calculating a tracking policy.

\subsubsection{Clustered Tracking Systems}

\begin{figure}
\begin{centering}
\includegraphics[width=0.9\columnwidth,trim=1cm 1cm 1cm 0.5cm]{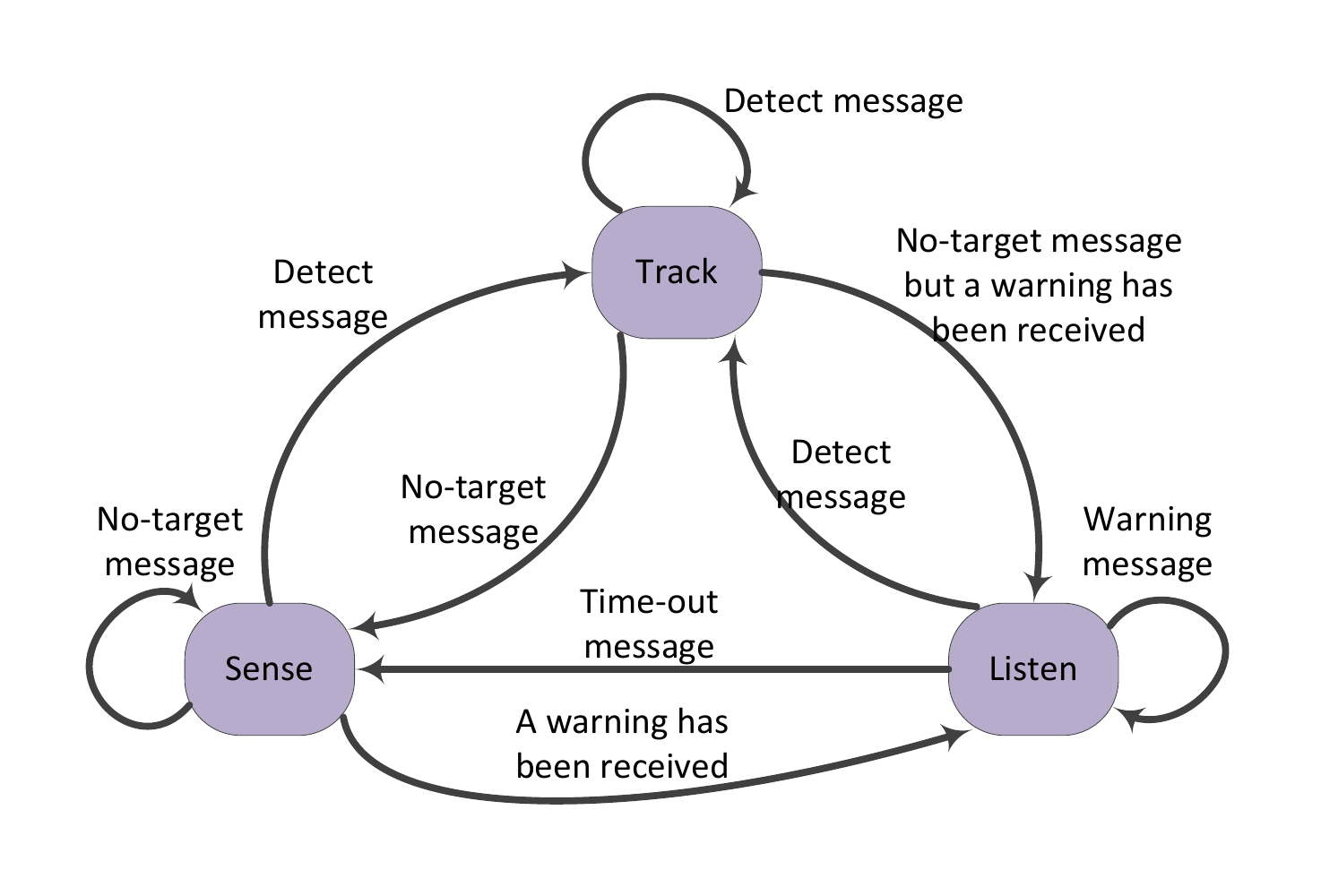}
\par\end{centering}

\caption{\label{fig:yeow2007energy} Sensor's state transition during target tracking as suggested by~\cite{yeow2007energy}. }
\end{figure}

In clustered architectures, object detection is performed by considering the resource availability at each device. Yeow \textit{et al.}~\cite{yeow2007energy} introduced the target tracking algorithm that considers both the spatial and temporal characteristics of sensor movement. The tracking problem is divided into two parts: (i) prediction of targets at lower level agents (LLAs), and (ii) activation management at a higher level agent (HLA). Here, the HLA is a cluster head that selects the set of active sensor nodes, i.e., LLAs. The algorithm is based on an HMDP model which minimizes the sensing rate of the sensors and maintains the detection accuracy. The model's states are shown in Figure~\ref{fig:yeow2007energy}. The cluster head operates under the states of periodic sensing, tracking, or active listening. In the periodic sensing state, the cluster head sleeps and wakes up periodically to sense any target. The tracking mode is activated when a target is detected. Finally, the listening mode is triggered when other cluster heads inform the detection of a target and it is expected to approach the cluster head covered area.

\begin{figure}
\begin{centering}
\includegraphics[width=0.6\columnwidth,trim=1cm 1cm 1cm 0.5cm]{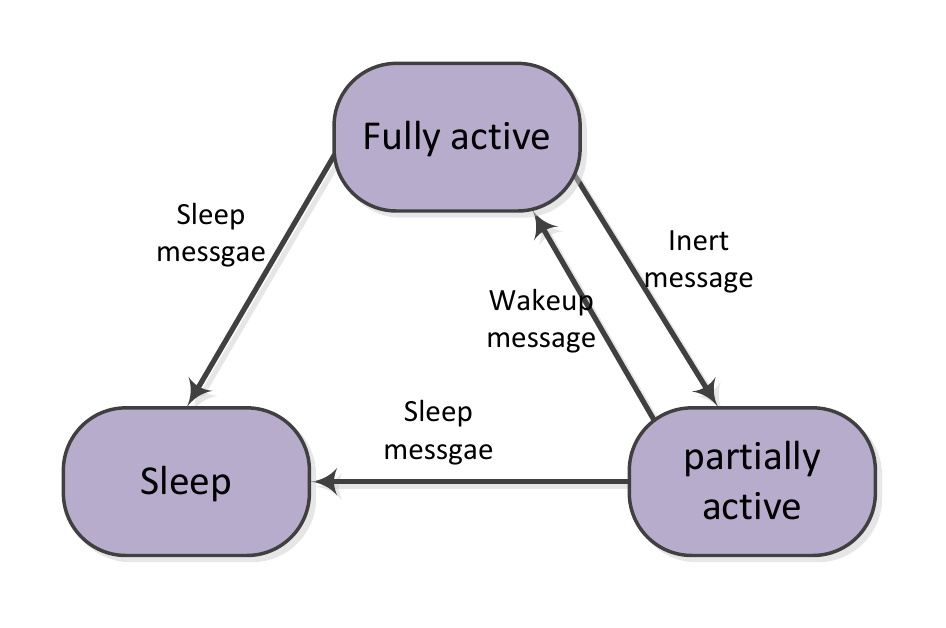}
\par\end{centering}

\caption{\label{fig:misra2012localized}State transition of object tracking sensors in surveillance systems as presented in~\cite{yeow2007energy}.}
\end{figure}

Misra and Singh~\cite{misra2012localized} considered the problem of precise object targeting in surveillance systems using WSNs. The minimum number of active nodes is selected by cluster heads to optimize energy consumption of the network. The node selection is optimized based on an MDP. A cluster head knows about an object's existence after receiving a message from neighboring clusters or when the object moves within the cluster head's detection area. The future object location is predicted using a Kalman Filter. Accordingly, a sensor node can be in sleep mode, partially active (sensing signals but not processing), or fully active mode (sensing and processing) as shown in Figure~\ref{fig:misra2012localized}. In the partially active mode, the cluster head can send a wake up request to the node to switch it to the full active mode.

\begin{figure}
\begin{centering}
\includegraphics[width=0.75\columnwidth,trim=1cm 1cm 1cm 0.5cm]{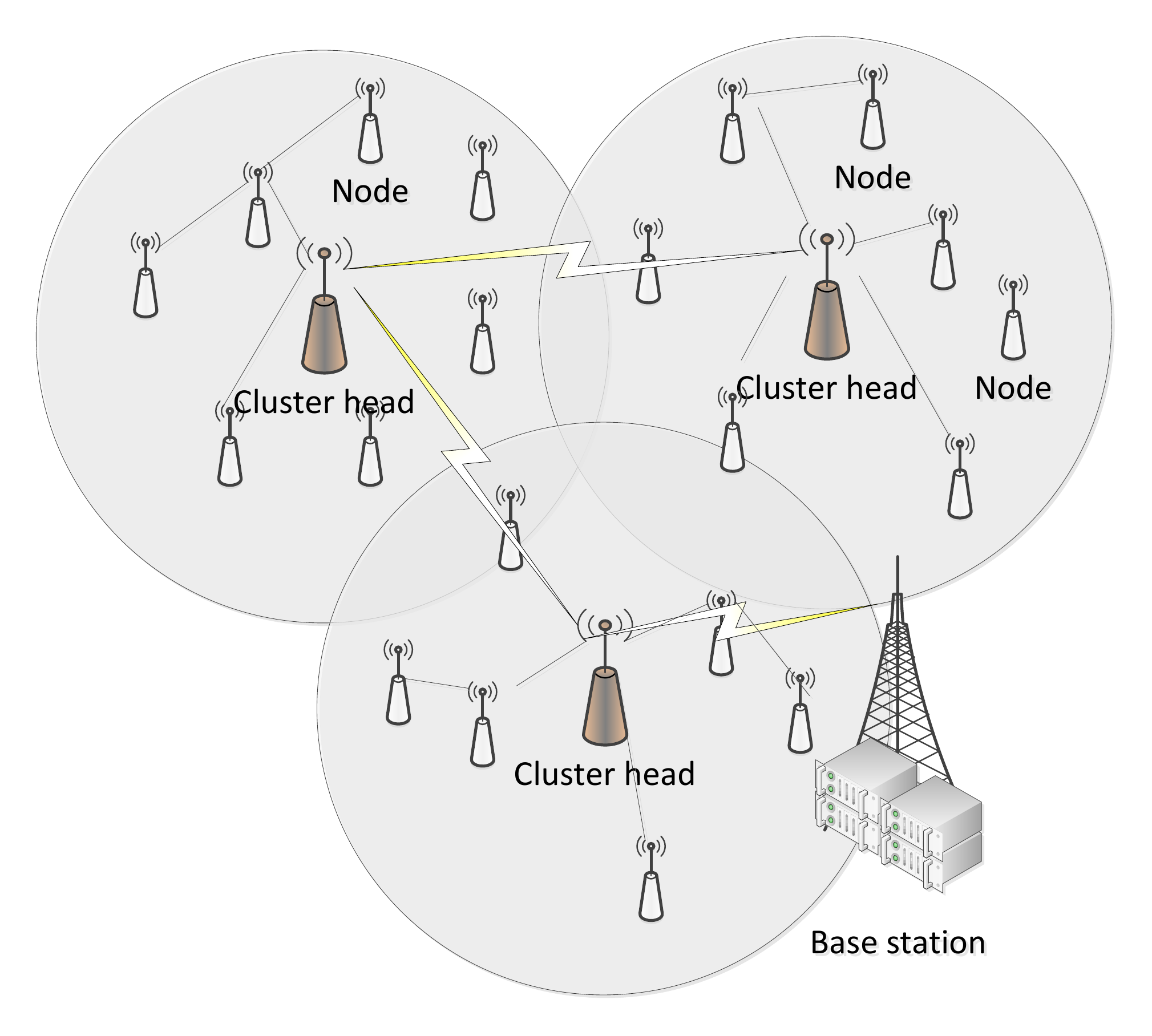}
\par\end{centering}

\caption{\label{fig:jamal2012event} A typical clustered architecture consisting of cluster heads and sensor nodes.}
\end{figure}

In cognitive radio, secondary users are allowed to opportunistically access the spectrum when it is not occupied by the primary users~\cite{akan2009cognitive}. Jamal \textit{et al.}~\cite{jamal2012event} used two CMDP models for efficient detection in cognitive radio WSNs. The system takes into account the detection accuracy, network congestion, and spectrum access constraints. The system is structured into two tiers. The upper tier consists of secondary users (cluster heads) to deliver messages to a base station. The lower tier comprises sensor nodes and the corresponding cluster head, i.e., a secondary user. A typical clustered architecture is shown in Figure~\ref{fig:jamal2012event}. The CMDP model is employed for balancing between high detection accuracy and low network congestion. Each node estimates the detection delay and sends it to the cluster head where a consensus delay decision is calculated. The second CMDP model is used to manage spectrum access at the upper tier by considering event arrival rates, queue status, link quality, service priority, and collision probability. At this upper tier, the actions comprise assigning the available spectrum to the secondary users.

\subsubsection{Multiple Target Tracking}

Li \textit{et al.}~\cite{li2009approximate} extended the model in~\cite{fuemmeler2008smart} to consider multiple target tracking. Again, the main goal is to analyze the tradeoff between energy saving through node sleep and detection performance. The centralized POMDP model uses a Monte Carlo method to find the belief states and to select a set of sensors for activation. The problem is solved using a combined method of particle filtering and a Q-value algorithm. In the same way, Zhang \textit{et al.}~\cite{zhang2013uts} presented a multiple target tracking solution based on a POMDP. The solution minimizes the number of active sensors to reduce the likelihood of sensor discovery (signal emission discovery) by enemy entities. Therefore, the balanced design is between the detection accuracy and the sensor's interception threat. The study assumes fixed sensors which operate independently. Each sensor can track a few targets simultaneously as long as the targets are within the detection range of the sensors. The POMDP's states correspond to target locations and moving velocities.

\subsubsection{Health and Body Wireless Sensor Networks}

Biometric sensors, e.g., pulse oximeters and electrocardiogram sensors (ECG), are widely used to detect human body activities such as in e-health applications. Au \textit{et al.}~\cite{au2011carer} discussed WSN-based chronic disease monitoring systems for real time tracking of human physical conditions. To prolong sensor lifespan, the scheduling algorithm is used to manage the sensor selection and activity by using the POMDP framework. In particular, the scheduling algorithm considers an equilibrium between detection accuracy and energy consumption by predicting if a sensor's activation is required in the next time instant. The state space contains the classified human's activities, and the sensors' readings are the observed beliefs. The action space includes the command for activating or switching off sensors. Similarly, Zois and Mitra~\cite{zois2013unified} introduced a model for activity detection in wireless body area networks (WBANs). The network is composed of heterogeneous nodes, and the node selection is optimized for maximum energy saving and maximum detection performance. Examples of detected human activities include standing up, running, walking, etc. Assuming noisy sensor outputs, a POMDP formulation is derived and solved using dynamic programming to obtain the selection strategy. The transition matrix is a square matrix that reflects the probabilities of switching between different body activities.

\begin{figure}
\begin{centering}
\includegraphics[width=0.65\columnwidth,trim=1cm 1cm 1cm 0.5cm]{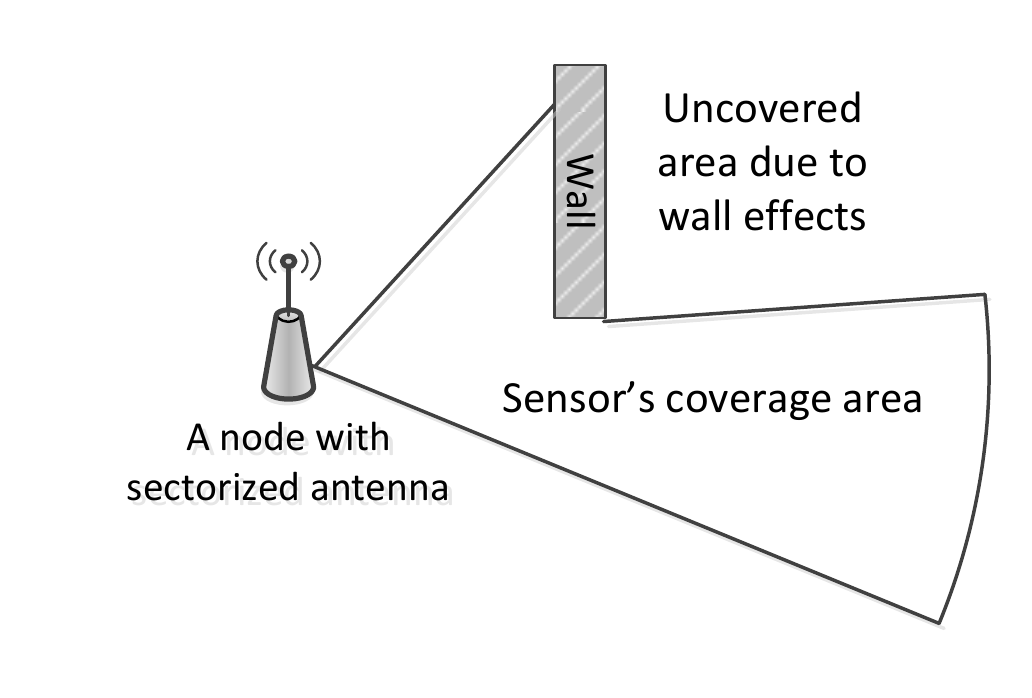}
\par\end{centering}

\caption{\label{fig:pietrabissa2013optimal}An example of poor coverage issue that generally occurs in indoor sensor applications~\cite{pietrabissa2013optimal}.}
\end{figure}

For fast asset transportation, Pietrabissa \textit{et al.}~\cite{pietrabissa2013optimal} discussed the tracking complication in hospitals including the localization of medical asset. This enables finding the moveable asset efficiently by using radio-frequency identification (RFID) technology. As an indoor application, the sensor coverage is affected by wall and equipment inside the building (Figure~\ref{fig:pietrabissa2013optimal}). The developed scheme also uses an agent to locate an asset and optimal path to bring the needed asset from storage location to asset's usage room. The states of the MDP correspond to the grid sectors of the hospital area, and the actions of the controller unit are movement operations to any of the four directions (north, south, east, or west). A reward is given if the transportation agent delivers the asset to the destination.

\subsubsection{Prioritized Data Delivery}

In~\cite{pino2014selective}, Pino-Povedano \textit{et al.} discussed the operation of selectively dropping unimportant data samples in target tracking applications. To maximizing the probability of delivering important messages over normal ones. In this application, an unimportant sample is that does not help in the object detection. The dropping scheme considers the node messages' importance, battery level, and transmission link cost. Each node takes an action of either sending or dropping the message to reduce its energy consumption over the radio transceiver based only on its local information. A successful delivery of important messages to the sink yields one unit of reward, and therefore a feedback is expected from the sink back to the source node. However, as the feedback may require long time to be received resulting in significant data load, the proposed scheme uses a suboptimal scheme based on two hop feedback, i.e., the outcome of data transmitted for two hop away from the source node. The simulation compares the suggested forwarding policy with a non-selective scheme that forwards all data samples. The proposed policy remarkably extends the network's lifetime and minimizes the total energy consumption.

The reviewed papers in this section have shown that the MDP models are useful for solving problems in sensing coverage and object detection. However, experiments using real world testbeds and deployments are still needed. Moreover, considering a grid topology is common in the literature, and hence further work is required for more general deployment distribution (e.g., the Poisson distribution). The next section reviews the adoption of MDPs for security and intrusion detection. The security component of a WSN ensures confidentiality and integrity of collected sensors' data~\cite{singh2011survey}.

\section{Security and Intrusion Detection}\label{sec:Security}

This section reviews the security related applications of an MDP in WSNs as summarized in Table~\ref{tab:security}. The few MDP-based security methods in the literature discuss the following issues: 
\begin{itemize} 
\item \emph{Intrusion detection}: One method for the detection of intrusion vulnerable node is based on an MDP. This is done by analyzing the correlation among samples collected from the nodes. Thus, the intrusion and intrusion-free samples are traced by an intrusion detection system (IDS).
\item \emph{Resource starvation attacks}: Resource starvation attacks aim at denying nodes from accessing the network resources such as wireless bandwidth. This is similar to the denial-of-service (DoS) attack. MDP-based security methods are developed to analyze the attacking entity behavior to select the optimal security configuration.
\end{itemize}

\begin{table*}
\caption{\label{tab:security}Summary of security surveyed security methods (SG = stochastic game, IDS = intrusion detection system, MTTF = mean time to failure, PDR = packet delivery ratio, RSSI = received signal strength indicator).}

\centering{}%
\begin{tabular}{|c|>{\centering}m{1.1cm}|>{\centering}m{1cm}|>{\centering}m{1.2cm}|>{\centering}m{3cm}|>{\centering}m{3cm}|>{\centering}m{2.6cm}|}
\hline 
\textbf{\noun{Security aspect}} & \textbf{\noun{Article}} & \textbf{\noun{Type}} & \textbf{\noun{Decision}} & \textbf{\noun{States}} & \textbf{\noun{Actions}} & \textbf{\noun{Rewards/costs}}\tabularnewline
\hline 
\hline 
\multirow{6}{*}{Intrusion detection} &~\cite{agah2004intrusion} & MDP & Centralized & Attacked sensor nodes  & Detect the intrusion's next attack & Prediction performance\tabularnewline
\cline{2-7} 
 &~\cite{alpcan2006intrusion} & SG & Centralized & Attack type & IDS: select a protecting action

Attacker: select an attack type & Attack detection\tabularnewline
\cline{2-7} 
 &~\cite{krakow2006control} & POMDP & Centralized & Intruder's location, sensor activations & Select active sensors & Detection performance\tabularnewline
\cline{2-7} 
 &~\cite{premkumar2008optimal} & MDP & Centralized & Sample, alarm & Control active nodes & False alarm, alarm delay\tabularnewline
\cline{2-7} 
 &~\cite{shen2012survivability} & SG & Centralized & Vulnerable, weak, risk,  & IDS: defend, do not defend

Attacker: attack, do not attack & MTTF\tabularnewline
\cline{2-7} 
 &~\cite{huang2013shielding} & MDP & Centralized & Node's state (under-attack or secure) & Defense a node & Intrusion detection\tabularnewline
\hline 
\multirow{3}{*}{Resource starvation attacks} &~\cite{mccune2005detection} & MDP & Centralized & Attacker's detection (detected or undetected) & Defense a node & Attack detection\tabularnewline
\cline{2-7} 
 &~\cite{li2012optimal} & MDP & Centralized & Channel jamming (PDR \& RSSI) & Activate an anti-jamming method & Energy, overhead, channel hopping cost\tabularnewline
\cline{2-7} 
 &~\cite{liu2014game} & SG & Centralized & Coordinator state (hacked or normal) & IDS: defend, do not defend

Attacker: attack, do not attack & Hop count\tabularnewline
\hline 
\end{tabular}
\end{table*}

\subsection{Intrusion Detection and Prevention Systems}

An intrusion detection system (IDS) monitors the nodes' collected data for abnormal samples. An abnormal reading is treated either as an indication of a malfunctioned sensor node or an external malicious attack. Agah \textit{et al.}~\cite{agah2004intrusion} addressed the problem of intrusion detection in WSNs by determining the most probable vulnerable nodes in the network. Thus, a vulnerable node can be protected and defended by further security mechanisms. The idea behind this design is to minimize the resource consumption in terms of memory and energy in protecting the network by restricting the number of protected nodes. One of the introduced mechanisms to define the vulnerable nodes is obtained from an MDP formulation. The MDP formulation is to predict the attacker's behavior and the next attacked node, i.e., the most vulnerable node. Then, the IDS receives the reward based on its prediction accuracy. That is, if the attacker attacks the protected node, this results in high reward value. The states of the MDP is the different nodes in the network and the attacker will move between these states. Additionally, the IDS will predict the transition probabilities between the states. The IDS receives a positive reward if it successfully predicts the next attacked node and a negative reward upon a failed prediction. 

Alpcan and Basar~\cite{alpcan2006intrusion} considered the problem of intrusion detection in WSNs using a 2-player zero-sum stochastic game. The IDS is the first player, aiming to protect the network. The second player is an attacking entity. The attacking entity takes actions by deciding an attack type that it can perform. The IDS action space includes passive and active action. Alarm activation is an example of passive actions, and collecting more information is an example of the active actions. The game state represents the detected attack at a time instant. Thereby, the transition matrix contains the probabilities of switching from one attack to another. As a zero-sum game, a successful IDS prediction of the attack results in a positive reward for the IDS and the negative reward for the attacker, and vice versa for a failed prediction by IDS.  

In order to minimize energy consumption of an IDS, Krakow \textit{et al.}~\cite{krakow2006control} considered the design of an energy efficient perimeter security system using WSNs and a POMDP. In particular, the POMDP model is to optimize the tradeoff between the detection performance and energy consumption by predicting the future location of an intruder. The system assumes partial information about the intruder state, and the posterior probabilities of the state beliefs are updated over time. The states consist of the intruder location and velocity, and the activation of other nodes. Then, the centralized POMDP policy predicts the activation decision for each sensor. Similarly, Premkumar and Kumar~\cite{premkumar2008optimal} suggested an energy efficient, MDP-based scheme for detecting intrusions using WSNs. During the system sampling state, a central unit coordinates all sensors into two operational subsets: an active and a sleep subset. The reward function takes into account the cost of false alarm, alarm delay, and collected samples using sensors. 

Shen \textit{et al.}~\cite{shen2012survivability} proposed a stochastic game-based attack detection mechanism for WSNs. The mechanism detects future attacks and the probabilities of changing the attack behaviour. Similar to~\cite{alpcan2006intrusion}, the problem is modeled as a 2-player zero-sum stochastic game. The mechanism maximizes the mean time to failure (MTTF) of nodes, which is a reliability metric. Therefore, an attacked node can be in one of three states: vulnerable, weak, and risk states. The attacker has two actions of whether to attack the nodes or not. The defending system takes protection actions, or it stays idle. The attacker receives a positive reward if it attacks the network while the protection system decides to stay idle. The simulation shows that the MTTF decreases as the attacking probabilities increase, and the survival lifetime is proportional to the number of nodes.

Furthermore, Huang \textit{et al.}~\cite{huang2013shielding} proposed an MDP-based intrusion detection and protection scheme for WSNs. The MDP framework detects a set of the vulnerable nodes to intrusion attacks at each time instant. The IDS coordinator receives a positive reward when it successfully predicts and secures the attacked nodes, and a negative reward if it fails to do so. The IDS stores the attackers' information and patterns, such as the time and interval of each attack, to predict future intrusion behavior and the time of their occurrence.

By contrast, the algorithms proposed in~\cite{agah2004intrusion,alpcan2006intrusion,krakow2006control,premkumar2008optimal,shen2012survivability,huang2013shielding} require an offline learning phase at a central unit. This centralized design incurs high costs for data gathering to a base station. 

\subsection{Resource Starvation Attacks}

Resource starvation attacks aim at stopping WSNs from normal operation by consuming network resources. For example, McCune \textit{et al.}~\cite{mccune2005detection} proposed a security mechanism to prevent packet denial attacks in broadcast protocols. In this type of attacks, the adversary prevents the network nodes from receiving the broadcast messages sent by the base station. The proposed mechanism relies on receiving acknowledgment messages (ACKs) from a randomly selected subset of nodes, thereby preventing acknowledgment implosion problems. Acknowledgment messages are received from each node in the network. Consequently, the failure to receive the broadcast message is assumed to be due to the adversarial attack, not a result of networking congestion. The proposed mechanism uses an MDP to model the attacker. The two attacker's states are a detected and an undetected states. The actions reflect the chosen node by the attacker for a denial attack, and hence the system will try to protect that vulnerable node.

Li \textit{et al.}~\cite{li2012optimal} tackled the problem of radio jamming in WSNs which causes low data exchange rates among sensor nodes. The proposed framework implements many state-of-the-art methods, and each method solves only a specific jamming case and no general solution can handle all jamming cases. Therefore, the suggested framework dynamically enables a suitable method for the existing jamming case based on the characteristics of jamming attacks. This MDP-based adaptive framework enables selecting the anti-jamming scheme without any node reprogramming. Applying an anti-jamming technique at a specific time is considered as an MDP's action. The action is chosen based on the cost of different anti-jamming technique and sensed channel conditions. The channel conditions depend on the jamming nodes' transmit power formulated as packet delivery ratio (PDR) and received signal strength indicator (RSSI). Additionally, the cost of different anti-jamming techniques are identified by power adjustment cost, error control overhead, and channel hopping and scanning costs. 

Liu \textit{et al.}~\cite{liu2014game} studied the security issues of using centralized coordinator to manage WSNs. Specifically, attacking the coordinator node can severely degrade the network performance and throughput. For example, a simple jamming attack near the coordinator can stop the data flow. Therefore, a coordination selection method is suggested to minimize hop counts from ordinary nodes to the coordinator as well as to protect the coordinator from malicious attack. The defending mechanism is based on stochastic games. The coordinator is the defending player and the malicious entity is the attacking player. The state space includes both normal and attacked states. The actions are attack and defend. Using the Network Simulator (NS-2) and a jamming attack scenario, it is shown that selecting a new un-attacked coordinator to manage the network topology increases the total throughput and lifetime of the network.

In summary, the existing literature of MDP-based security methods is relatively small. Clearly, stochastic games are well suited for probabilistic security monitoring and attack remediation, and further research is required to expand the preliminary results reviewed in this section. By contrast, using fully observable MDPs for preventing channel jamming seems to be practical because they do not require high computational resources. The following section is dedicated to custom applications of WSNs that have been addressed using MDP-based algorithms. Each of these applications comes with special requirements in terms of sensor types, energy consumption, and design objectives.

\section{Custom Applications and Systems}\label{sec:Custom-applications}

This section describes many WSN applications that have been enhanced using the MDP framework including visual and modeling systems, traffic management and road safety, unattended wireless sensor networks (UWSNs), agriculture wireless sensor networks (AWSNs), wireless sensor and vehicular networks (WSVNs). 

\subsection{Visual and Modeling Systems}

Zhuang\textit{ et al.}~\cite{zhuang2005power} addressed the combination of Web services for real time data retrieval and search in WSNs, e.g., for equipment monitoring applications. In this context, a Web service provides an efficient mechanism to deliver the physical data for many applications in a uniform manner, and hence it provides an interoperable data exchange. The continuous and massive data collected by sensor nodes requires an optimized query architecture. The raw sensor data is represented using the Extensible Markup Language (XML) which facilitates data processing and information retrieval. This design adopts an MDP to estimate the uncertainty in query results. The states include the service's stateful resources (i.e., sensors with temporal data) which can be queried by exchanging messages among web services.

Many recent applications of WSNs are based on camera sensors which require special resource management in terms of energy and bandwidth resources. Therefore, Fallahi \textit{et al.}~\cite{fallahi2009dynamic} discussed the assurance of quality of service (QoS) in WSNs consisting of video camera nodes that capture and send video to a fusion center. In addition to energy limitations because of sending large size data, the QoS provisioning imposes another constraint. The authors therefore proposed an MDP-based scheme for adaptive power allocation while considering the scene generation rate, transmission buffer allocation, and physical channel parameters. The MDP formulation considers the moving picture experts group (MPEG) coded video, and an optimal policy is found using dynamic programming. The considered QoS metrics are the energy saving, data dropping rate, and transmission delay.

\subsection{Traffic Management and Road Safety}

In~\cite{witkoskie2006random}, Witkoskie \textit{et al.} considered the problem of multiple target road monitoring systems fixed at road intersections. An MDP resource management algorithm is developed to manage the sensor activation. The road is divided into monitoring segments and a unified hypothesis about any hostile existence is built by considering sensors' outputs at each road segment. The states represent the system knowledge about the number of discovered targets. Therefore, if the system state, i.e., knowledge, about the targets is high, less samples are needed from the sensors, and more sensors can be switched to sleep mode to save energy.

\subsection{Unattended Wireless Sensor Networks}

Unattended wireless sensor networks (UWSNs) are designed to work for relatively long time without maintenance or battery change. Accordingly, Misra and Ankur~\cite{misra2011policy} presented an energy saving scheme for selective node activation in UWSNs. The scheme considers the energy consumption, topology maintenance, and reliability requirements in the MDP formulation. The scheme considers the distance between nodes, node's energy budget, and number of neighboring nodes. A global positioning system (GPS) device is assumed to be available at each node. The node transits between five states: sleep, active, neighbor discovery, emergency detection, and idle (no sensing) modeS. Likewise, Ghataoura \textit{et al.}~\cite{ghataoura2011autonomic} investigated the use of UWSNs in monitoring and security applications. A POMDP is used to extract the temporal context of the threat and determine the optimized transmission time.

Self-management solutions enable nodes to reconfigure themselves if they experience software and hardware failures. For example, Bhuiyan \textit{et al.}~\cite{bhuiyan2013local} discussed WSN maintenance in event detection applications by proposing a local maintenance and failure monitoring routine. Specifically, the suggested maintenance algorithm detects specific network failures that can occur during event monitoring, such as link and node faults. Accordingly, the algorithm activates a prompt maintenance action. The node autonomously detects its faults using an SMDP during its active mode. The active mode includes three states: pre-processing, running and idle modes. The node is considered to be failed if the current state is inconsistently modified, i.e., does not follow the transition matrix. Moreover, the study considers link faults and suggests an election scheme for the link monitoring coordinator that uses a Markov chain in its link estimation process.

\subsection{Agriculture Wireless Sensor Networks}

Shuman \textit{et al.}~\cite{shuman2010measurement} developed an energy efficient soil moisture sensing scheme using a POMDP. The scheme schedules the sampling task of the sensors in such a way that sparse samples are taken for the area of interest. Then, the nodes are assumed to be noiseless and operate in one of two modes: active or sleep modes. The actions correspond to sensing moisture measurements at different soil depths. These assumptions are used to cast the POMDP problem into an infinite time horizon MDP structure which can be solved by dynamic programming. Similarly, Wu \textit{et al.}~\cite{wu2012situ} studied soil moisture sensing using a few readings. The measurement management scheme is designed based on a POMDP. The locations of measurements over time are considered as the states, and the action space describes whether sensing the moisture is required at each state. The moisture values are assumed to be quantized to a finite number of states. The proposed scheme is compared with an open-loop method that is based on compressive sensing. The POMDP method is more precise and achieves a balanced tradeoff between the sensing cost and recovery error. However, the compressive sensing method is less computationally intensive and does not require statistical knowledge of the underlying random process.

\subsection{Wireless Sensor and Vehicular Networks (WSVNs)}

\begin{figure}
\begin{centering}
\includegraphics[width=1\columnwidth]{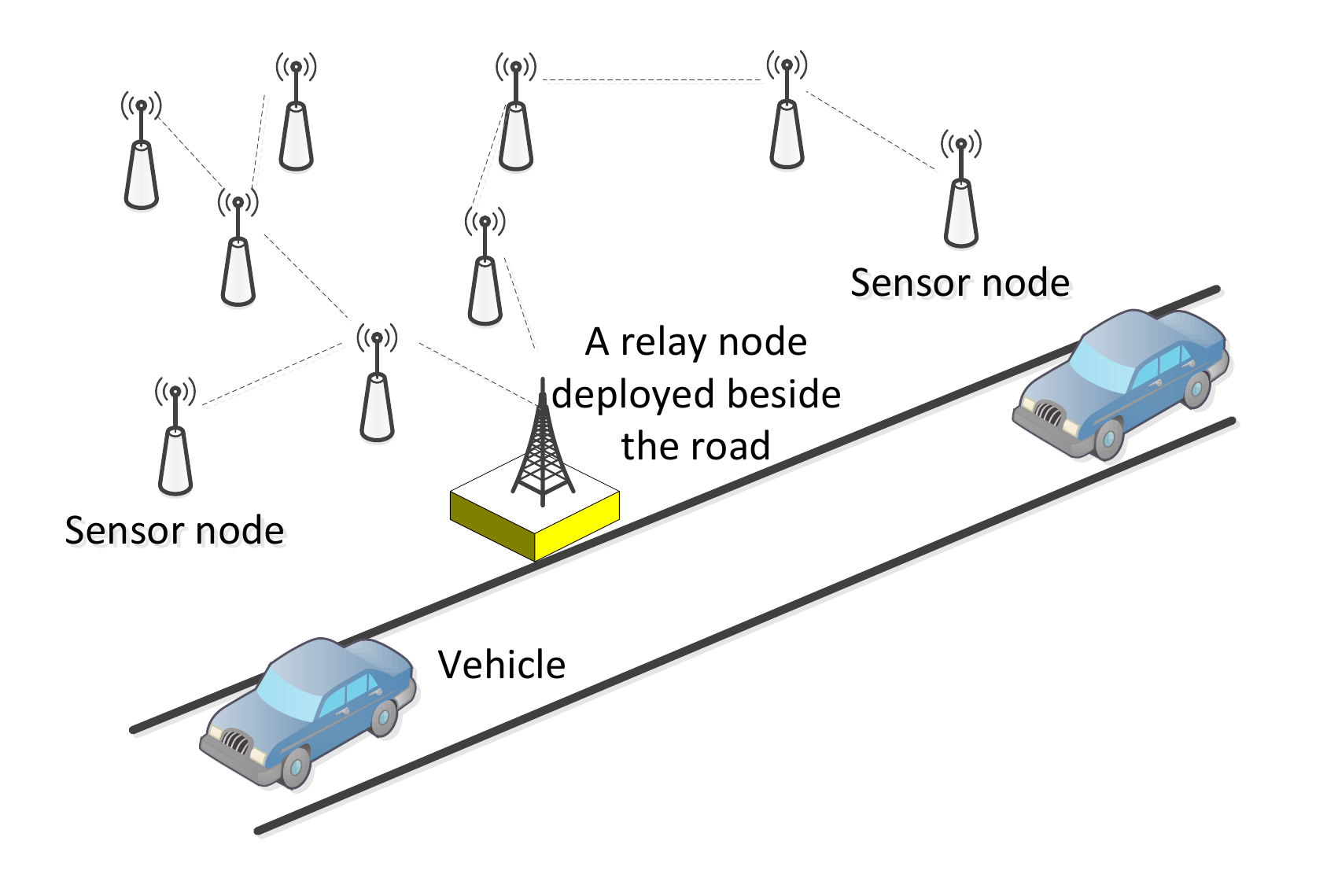}
\par\end{centering}

\caption{\label{fig:wsvn}An example of wireless sensor and vehicular networks (WSVNs).}
\end{figure}

Wireless sensor and vehicular networks (WSVNs) use moving vehicles such as cars and buses to collect data from deployed sensors and then deliver the data to the base station. An example of WSVNs is shown in Figure~\ref{fig:wsvn}. In~\cite{arshad2012fair}, Arshad \textit{et al.} studied the buffer allocation problem of vehicular nodes in sparsely deployed WSNs. The proposed scheme provides fair service to all source nodes that are selected to transmit their data through the roadside relay node by managing their buffer requirements. An SMDP model is developed to provide a look-up table of the optimal data collection decision at the relay node. The buffer size of the relay node is divided into multiple levels, and the current state depends on the buffer occupancy with sensor data. At each time instant, a relay node decides to receive the nearby sensor's data, drop data, or keep the current state until the data in the buffer is delivered to a passing vehicle. The study assumes that the data sensing process follows a Poisson distribution, and the buffer state duration is independent and identically distributed (iid).  

Similarly, Choi \textit{et al.}~\cite{choi2013delay} designed an MDP model to optimize data routing in WSVNs. The problem formulation takes into account the data delivery delay which is affected by the vehicle's speed and distribution. The state space consists of the data delivery at the intersections. The data delivery depends on the link condition which is decided using the MDP model.

\section{Future trends and open issues}\label{sec:Future-research}

WSNs find new applications and serves as a key platform in many smart technologies and Internet of Things (IoT). This continually introduces open design challenges in which MDPs can be used for making decisions. In this section, we discuss a few open research problems that have not been fully studied in the literature, and they require further research attention. These future research directions are discussed under three topics: (i) challenges of applying MDPs to WSNs, (ii)~emerging MDP models, and (iii) emerging topics in WSNs.

\subsection{Challenges of Applying MDPs to WSNs}

The MDP framework is a powerful analytical tool to address stochastic optimization problems. The MDP framework has proven its applicability in many real world applications such as finance, agriculture, sports, etc~\cite{White_1993_A_survey,Feinberg_handbook_2002_Handbook_of,Sigaud_book_2010_Markov_Decision,Qiying_book_2008_Markov_Decision,Xianping_book_2008_Introduction_to}. However, there are still some limitations that need further research study. 

\subsubsection{Time Synchronization}

Most existing studies assume perfect time synchronization among nodes. This assumption enables the network nodes to construct a unified MDP cycle (sense current state, make decision and take actions, sense new state, etc). Therefore, the clock of the node must be adjusted to a central timing device (see~\cite{sichitiu2003simple,elson2003time} for time synchronization algorithms in WSNs). Besides, the clock may not be perfectly synchronized because of various delay. The mechanisms to address these issues must be developed.

\subsubsection{The Curse of Dimensionality}

This is an inherent problem of MDPs when the state space and/or the action space become large. Consequently, we cannot solve MDPs directly by applying standard solution methods. Instead, approximate solutions~\cite{Sutton_book_1998_Introduction_to,Warren_book_2011_Approximate_Dynamic,Farias_2003_The_linear} are usually used. The work in \cite{chobsri2009parametric,pandana2005near,li2012analytical,aprem2013transmit} present some examples of using approximate solutions to reduce the complexity of MDP-based methods in WSNs. 

\subsubsection{Stationarity and Time-Varying Models}

It is assumed that the MDP's transition probabilities and reward function are time invariable. Nevertheless, in some systems, this assumption may be infeasible. There are two general methods to deal with non-stationary transition probabilities in Markov decision problems. In the first solution, an online learning algorithm, e.g.,~\cite{Jia_2009_online,Gergelyphdthesis}, is used to update the state transition probabilities and the reward function based on the environment changes.

In the second approach, the state space is extended by including time to deal with non-stationary transition probabilities. This idea derives from the fact that transition probabilities can be defined as a function of time. Thus, by using time as a state, the transition probabilities become stationary with state space. For example, conjugate MDPs (CoMDPs)~\cite{thomas2011conjugate} include selecting time-varying parameters when transiting from a current state to a next state. Examples of time-varying parameters include approximation weights and learning rates. After moving to a new state, the time-varying parameters are also updated. Therefore, a coordinate ascent method is used for the policy and time-varying parameter optimization. A related idea is found in the one-counter MDP (OC-MDP) model~\cite{brazdil2010one} which extends a basic MDP formulation by introducing a counter variable that is modified during state transition. In particular, the transition depends not only on the current state but also on the counter value. OC-MDPs include two types of states: random and controlled states. The transition of the random state is decided over a probability distribution. Alternatively, the transition from the controlled state is determined by a controller.

\subsection{Emerging MDP Models}

Recently, many new models and solution techniques have been introduced for MDPs. These recent advances can help in developing more effective WSN solutions and overcoming limitations of classical MDP-based models. Examples of these advances are summarized as follows.

\subsubsection{State Abstraction Using Self-Organizing Maps (SOMs)}

\begin{figure}
\begin{centering}
\includegraphics[width=0.85\columnwidth,trim=1cm 3cm 1cm 0.5cm]{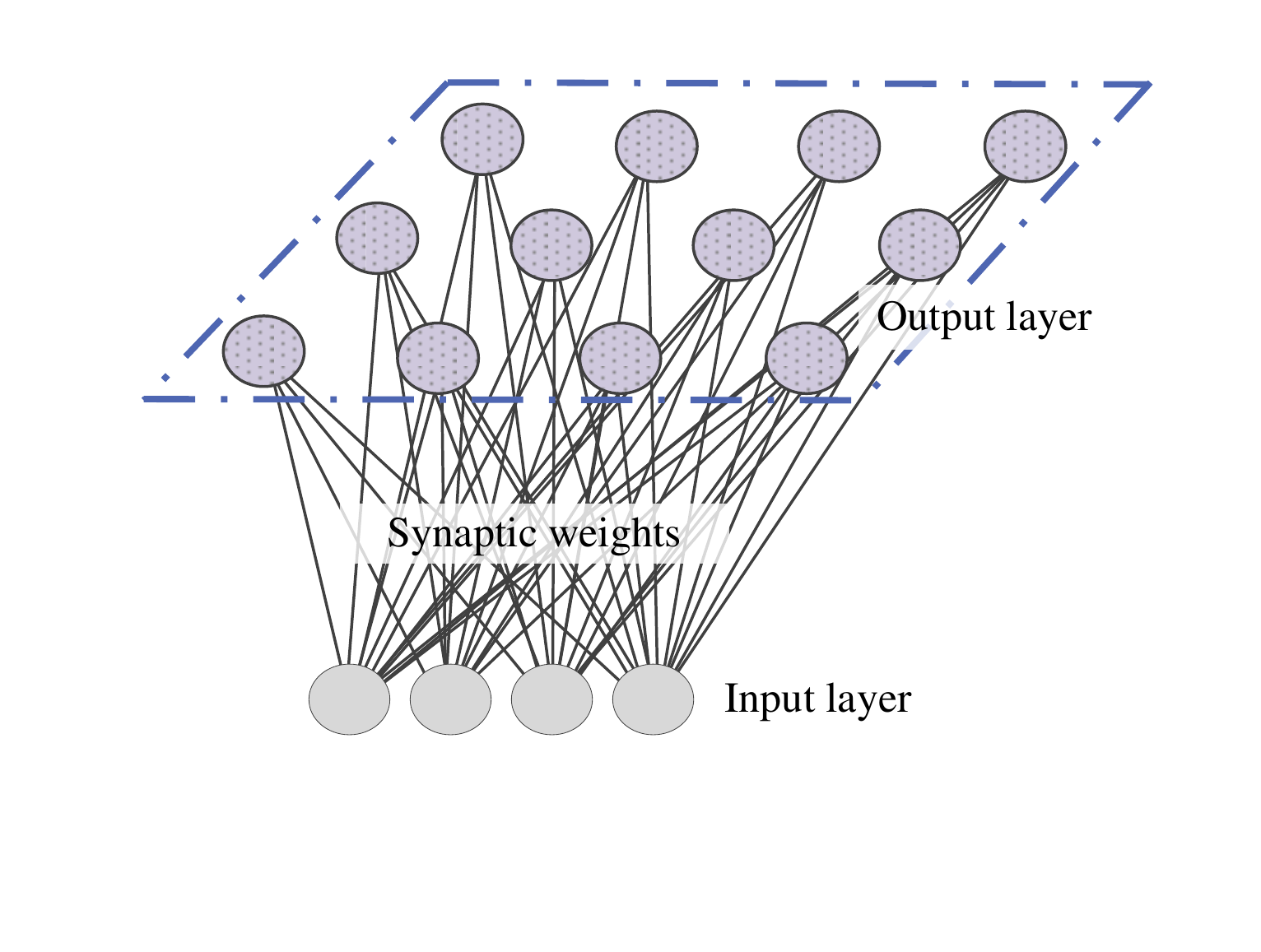}
\par\end{centering}

\caption{\label{fig:SOMs}A self organizing map with a 4D input space that is mapped to 12 distinctive classes in a 2D output lattice. Synaptic connections are tuned using an offline competitive learning over historical data.} 
\end{figure}

Self-organizing maps (SOMs)~\cite{kohonen1990self} classify continuous value sensory inputs into distinctive output classes. SOMs are unsupervised artificial neural networks that can learn high-level features from a historical data as shown in Figure~\ref{fig:SOMs}. For MDP state abstraction, the input layer is fed with the state parameters, and the high-level states are produced at the output layer. Thus, the generated states present the correlations between input parameters. It has been shown that using SOMs can automate the formulation of distinctive states for MDPs in general robotics~\cite{toussaint2004learning,provost2006developing}. Even though SOMs were used in a few applications~\cite{alsheikh2014machine}, the use of SOMs for MDP state formulation in WSNs is not well explored. Such exploration can reduce the complexity of solving problems with continuous and discrete state values which is a promising benefit for practical applications of MDPs in WSNs.

\subsubsection{Learning Unknown Parameters}

In a MDP framework, we assume that the transition probability function and the reward function are known in advance. In some application contexts, this requirement may be impossible. Therefore, learning algorithms~\cite{Sutton_book_1998_Introduction_to,Sigaud_book_2010_Markov_Decision} are used. Another direction is using robust MDPs (RMDPs)~\cite{wiesemann2013robust} that deal with the uncertainty in selecting modeling parameters (e.g., transition kernel) by learning these unknown parameters from historical data. An RMDP model is suitable for the systems where the long term expected reward is sensitive to the difference between the estimated and actual transition probabilities. This model provides a probabilistic confidence on the system performance and under worst-case conditions.

\subsubsection{Near-Optimal Solutions}

Sensor nodes are independent controllers located in an environment and their decisions have mutual effects on each other. Many Markov models were used to for multiple controllers as reviewed in this paper including multi-agent MDPs, distributed MDPs, and stochastic games. Nevertheless, most of the existing solutions assume that the nodes can observe the state of each other by exchanging information or through a central coordinator. This assumption may be inapplicable in some practical contexts because of noise, constrained-hardware, and battery limitation. Consequently, we have to consider other kinds of the Markov models, e.g., partially observable multi-agent MDPs. Two major candidates for such models are decentralized partially observable MDPs (DEC-POMDPs)~\cite{Bernstein_2002_complexity} and partially observable stochastic games (POSGs)~\cite{Hansen_2004_Dynamic}. Although these models formulate problems with partial observations and multiple controllers, their solutions are very complicated as explained in Section~\ref{sec:mdp_complexity}. Therefore, this leads to implementation difficulties in WSNs. Alternatively, a possible research direction is to derive near-optimal solutions and estimations for these methods, which incur less complexity.

\subsection{Emerging Topics in WSNs}

This section discusses three potential research opportunities for using MDPs in WSNs.
  
\subsubsection{Cross-Layer Optimized Sensor Networks (CLOSNs)}

The cross-layer optimization has been proposed to circumvent the limitations because of standard layer-based protocol design, and it is recently adopted in WSNs. A cross-layer architecture enables the interaction of protocols at different layers and supports multiple QoS objectives such as end-to-end (E2E) delay, bandwidth usage, loss rate, etc. This provides more flexibility to solve many issues in WSNs~\cite{mendes2011survey}. MDPs are suitable for optimizing multiple objectives at different layers, and a few works in the literature presented MDP-based cross-layer algorithms such as in data aggregation~\cite{lin2011autonomic}, transmission scheduling~\cite{xiong2010cross}, and object tracking~\cite{liu2006cross}. Accordingly, further research is required for a viable and universal design, and where the MDP model  can be used for the multi-objective optimization (e.g., resource allocation algorithms, distributed source coding, cross-layer signaling, secure transmission, etc).

\subsubsection{Cognitive Radio Sensor Networks (CRSNs)}

Cognitive radios are developed for efficient dynamic spectrum sharing. CRSNs benefit from dynamic spectrum access, and they can be applied to many applications such as indoor and heterogeneous sensing, multimedia networks, and real-time surveillance applications~\cite{akan2009cognitive}. A few works in the literature discussed the potentials of using MDPs in CRSNs with a centralized coordinator, e.g.,~\cite{park2012optimal,jamal2012event}. However, there are many further research potentials for using MDPs in CRSNs including QoS aware routing methods, distributed spectrum sensing, and opportunistic data collection and transmission. Moreover, game-theoretic studies for CRSNs are interesting research directions where nodes independently and rationally take spectrum access actions. A stochastic game approach enables finding any kind of equilibrium solutions and minimizing interference among transmissions of competing nodes. On the complexity aspect, finding optimal MDP solutions in CRSNs depends on the number of sensor nodes, and therefore exploring suboptimal and estimation solution with less complexity is important for large scale CRSNs.

\subsubsection{Privacy Models}

WSNs are finding more applications in human-centric services, and hence the collection of private and confidential data becomes a crucial issue. Privacy is required to protect data form suspicious entities. For example, many studies discussed the patients' privacy concerns when using a wireless body area network to gather data about daily health conditions~\cite{li2010data,al2012security}. However, the resource limitations of WSNs impede the wide inclusion of privacy solutions to protect message confidentiality~\cite{ortolani2011events}. MDPs can be used to find a balanced tradeoff between the complexity of privacy models and energy consumption. Furthermore, another direction is to use stochastic games to model the interaction between a WSN and malicious entities.

\subsubsection{Internet-of-Things (IoT)} 
The IoT consists of sensing devices and benefits from the Internet infrastructure, and hence the WSN technology is a key component of many IoT applications. Herein, sensor nodes (referred to as smart objects) require energy-efficient solutions and interact with a variety of computing systems. An MDP is a promising tool to optimize the multi-objective optimization in IoT systems. For example, Li \textit{et al.}~\cite{li2012modeling} studied the integration of web services in IoT systems while considering the reliability and resource consumption (e.g., energy and bandwidth cost) using an MDP model. Yau \textit{et al.}~\cite{yau2014intelligent} proposed an MDP-based intelligent planning in mobile IoT that incorporates mobile cloud systems into the standard IoT technology. In 2020, 24~billion devices are expected to be interconnected~\cite{gubbi2013internet}. Therefore, an important research direction is proposing scalable and distributed MDP solutions for decision making in IoT systems.

\section{Summary}\label{sec:Summary}

This paper has provided the extensive literature review related to a Markov decision process framework and its applications in wireless sensor networks. An introduction to the Markov decision process has been given, and important extension models have been also reviewed. Then, many design of the Markov decision process in wireless sensor networks have been discussed including data exchange and topology formation, resource and power optimization, area coverage and event tracking solutions, and security and intrusion detection methods. Finally, the paper has discussed about a few interesting research directions.


\section*{Acknowledgements}

This work was supported in part by Singapore MOE Tier 1 grants (RG18/13 and RG33/12).

\bibliographystyle{IEEEtran}

\bibliography{Bibtex/Applications,Bibtex/Data_aggregation_routing,Bibtex/Dynamic_optimization_resource_allocation,Bibtex/Energy_management_harvesting,Bibtex/Relay_selection_placement,Bibtex/Scheduling_mac,Bibtex/Security_intrusion_detection,Bibtex/Sensor_coverage_phenomenon_monitoring,Bibtex/solving_mdp,Bibtex/Target_tracking_localization,Bibtex/Transmission_strategy_neighbor_discovery,Bibtex/Others,Bibtex/Future_trends}

\end{document}